\documentclass[showpacs,preprintnumbers,pra,twocolumn]{revtex4-1}

\usepackage{graphicx}
\usepackage{amsmath}
\usepackage{times}

\newcommand{\therm}{\mathrm{th}}
\newcommand{\ext}{\mathrm{ext}}
\newcommand{\innt}{\mathrm{int}}
\newcommand{\imf}{\mathrm{if}}
\newcommand{\cool}{\mathrm{2}}

\begin{document}

\title[Heterodyne photodetection measurements on cavity optomechanical systems:\\ Interpretation of sideband asymmetry and limits to a classical explanation]{Heterodyne photodetection measurements on cavity optomechanical systems:\\ Interpretation of sideband asymmetry and limits to a classical explanation}
\author{Kjetil B{\o}rkje}
\affiliation{Royal Norwegian Naval Academy, Postboks 83, Haakonsvern
NO-5886 Bergen, Norway}

\date{\today}

\begin{abstract}
We consider a system where an optical cavity mode is parametrically coupled to a mechanical oscillator. A laser beam driving the cavity at its resonance frequency will acquire red- and blue-shifted sidebands due to noise in the position of the mechanical oscillator. In a classical theory without noise in the electromagnetic field, the powers of these sidebands are of equal magnitude. In a quantum theory, however, an asymmetry between the sidebands can be resolved when the oscillator's average number of vibrational excitations (phonons) becomes small, i.e., comparable to 1. We discuss the interpretation of this sideband asymmetry in a heterodyne photodetection measurement scheme and show that it depends on the choice of detector model. In the optical regime, standard photodetection theory leads to a photocurrent noise spectrum given by normal and time ordered expectation values. The sideband asymmetry is in that case a direct reflection of the quantum asymmetry of the position noise spectrum of the mechanical oscillator. Conversely, for a detector that measures symmetric, non-ordered expectation values, we show that the sideband asymmetry can be traced back to quantum optomechanical interference terms. This ambiguity in interpretation applies not only to mechanical oscillators, but to any degree of freedom that couples linearly to noise in the electromagnetic field. Finally, we also compare the quantum theory to a fully classical model, where sideband asymmetry can arise from classical optomechanical interference terms. We show that, due to the oscillator's lack of zero point motion in a classical theory, the sidebands in the photocurrent spectrum differ qualitatively from those of a quantum theory at sufficiently low temperatures. We discuss the observable consequences of this deviaton between classical and quantum theories.
\end{abstract}


\maketitle

\section{Introduction}

In canonical cavity optomechanics, the frequency of an optical cavity mode is linearly dependent on the position of a mechanical oscillator. Light trapped in the optical cavity can then be used not only to measure the position of the mechanical oscillator, but also to influence the oscillator's motion. This turns out to be a very useful tool for studying large scale mechanical systems in the quantum regime \cite{Aspelmeyer2014RMP,Poot2012PhysRep}. Similar physics can be realized in electromechanical systems where microwave resonators in superconducting circuits take the role of optical cavities and lasers are replaced by signal generators. The study of opto- and electromechanical systems can for example lead to technological advances within sensing or signal processing \cite{Metcalfe2014ApplPhysRev}, but it is also a promising route towards confirming or ruling out deviations from standard quantum theory \cite{Adler2009Science,Arndt2014NatPhys}.

The motion of a mechanical oscillator is typically dominated by thermal noise even in a cryogenic environment, unless the oscillator's resonance frequency is very large. To remove thermal excitations, a useful tool is the technique of cavity-assisted sideband laser cooling. Beyond the atomic scale \cite{Diedrich1989PRL}, this technique was first used, both in electromechanical \cite{Teufel2011Nature} and optomechanical \cite{Chan2011Nature} systems, to cool a particular mechanical mode of micrometer scale structures close to the ground state. Recently, similar results have been achieved with more massive systems such as thin dielectric membranes with transverse dimensions on the millimeter scale \cite{Underwood2015PRA,Purdy2015PRA,Peterson2016PRL}.

In addition to cooling, light or microwaves can also be used to confirm that a mechanical oscillator has in fact been cooled to the quantum regime. To see this, let us imagine that a probe laser beam is sent into the optical cavity. The motion of the mechanical oscillator leads to sidebands in the light leaking out of the cavity. The sideband frequencies are one mechanical frequency above and below the probe beam's frequency. Let us further assume that the probe beam frequency is equal to the cavity mode's resonance frequency. In a fully classical theory in which the electromagnetic field and the laser has no noise, these sidebands are then of equal strength. However, in a quantum theory \cite{Marquardt2007PRL,Wilson-Rae2007PRL}, the ratio between the fluxes of blue- and red-shifted light is given by $\tilde{n}_\therm/(\tilde{n}_\therm + 1)$, where $\tilde{n}_\therm$ is the mechanical oscillator's average number of excitations (i.e., phonons). This sideband asymmetry can thus be used to determine the mechanical oscillator's effective temperature. 

It can be challenging to filter out these sidebands from the carrier frequency in order to measure the individual sidebands. However, this is not necessary, as the technique of heterodyne detection can be used to address the individual sidebands in the Fourier domain. Khalili et~al.~\cite{Khalili2012PRA} and Weinstein et~al.~\cite{Weinstein2014PRX} have discussed the interpretation of sideband asymmetry measured using this technique. In these works, it was claimed that sideband asymmetry in heterodyne detection originates from the oscillator's response to quantum noise in the electromagnetic field, or in other words, from correlations between quantum noise and the mechanical oscillator position. This was contrasted with measurements of sideband asymmetry by direct photodetection of the filtered sidebands, in which the asymmetry can be traced back to the mechanical oscillator's intrinsic quantum noise \cite{Weinstein2014PRX}. Nevertheless, in standard quantum theory, one finds the magnitude of the sidebands and their asymmetry to be the same in both interpretations. One could thus argue that this issue is of no scientific interest as the two interpretations cannot be distinguished experimentally, although this assumes {\it a priori} that standard quantum theory is correct. 

The fact that sideband asymmetry can arise from the oscillator's response to noise in the electromagnetic field is perhaps a cause for concern, since asymmetry can originate not only from quantum noise, but also from classical noise \cite{Jayich2012NJP,Safavi-Naeini2013NJP}. Classical noise in the electromagnetic field can sometimes be ruled out with the use of sufficiently high quality detectors and by filtering of laser noise. However, if an accurate noise characterization cannot be made, one might worry that sideband asymmetry cannot be interpreted as a signature of the quantum nature of the mechanical oscillator. It should be mentioned that other techniques for measuring an oscillator's quantum zero point motion are possible, for example by introducing a nonlinear resource \cite{Lecocq2015NatPhys}. That being said, sideband thermometry of the kind we have discussed is likely to be the first and easiest choice in many setups and will probably be extensively used also in future experiments. A detailed study of this measurement technique is therefore in order.

In this article, we first examine the interpretation of sideband asymmetry with heterodyne detection in a bit more detail than in Refs.~\cite{Khalili2012PRA,Weinstein2014PRX}. We will show that the interpretation of the asymmetry depends on the model used for the detector, and thus it requires a detailed knowledge of the measurement process. We study two different detector models for calculating the heterodyne spectrum, one defined from symmetrized expectation values and one defined from normal and time ordered expectation values. The difference in interpretation is closely related to the older discussion of whether photocurrent shot noise is a result of the photodetection process itself or whether it comes from quantum noise in the electromagnetic field. Carmichael gave an illuminating discussion of this issue in Ref.~\cite{Carmichael1987JOptSocAmB}. Spontaneous emission provides another example of how interpretation can depend on operator ordering \cite{Milloni1984AmJPhys}. 

Specifically, we study the optical regime where heterodyne detection is performed by linearly combining the light from the cavity with a detuned local oscillator, for example by using a beam splitter, before detection in a photomultiplier. In this case and in the absence of classical electromagnetic noise sources, we will show that it follows from standard photodetection theory that the sideband asymmetry in the photocurrent spectral density can be explicitly expressed in terms of the quantum asymmetry of the noise spectrum of the mechanical oscillator's position. In fact, unlike in the analyses of Refs.~\cite{Khalili2012PRA,Weinstein2014PRX}, correlations between quantum noise and mechanical oscillator position cannot contribute to the photocurrent spectrum in this setup, since a photomultiplier cannot detect quantum vacuum noise. We note that this result is not specific to mechanical oscillators, but applies to any degree of freedom that couples linearly to vacuum fluctuations of the electromagnetic field.

We also address the question of whether a classical interpretation of the sideband asymmetry is always possible in cases where classical noise in the electromagnetic field cannot be ruled out. We find that, while a classical model that gives the right amount of asymmetry can always be constructed, the sidebands themselves are qualitatively different in quantum and classical theories below a certain temperature. In a quantum theory, the height of the blue sideband is proportional to $\tilde{n}_\therm$ and hence always positive. On the other hand, in a classical theory where the oscillator does not have zero point fluctuations, the blue sideband height can become negative, i.e., the sideband peak turns into a dip. We discuss observable consequences of this that can potentially be used to rule out a classical theory. However, we also point out that since classical electromagnetic field noise sets a lower limit on the effective temperature of the mechanical mode that is attainable by laser cooling, the deviation between classical and quantum theories would only be observable if the oscillator is cooled by other means of cooling than laser cooling.

The detection method we have described and that we will analyze below have been used in the experiments reported in Refs.~\cite{Brahms2012PRL,Jayich2012NJP,Underwood2015PRA,Purdy2015PRA,Peterson2016PRL}. An alternative method consists of measuring one sideband at a time with different probe beam frequencies \cite{Weinstein2014PRX,Safavi-Naeini2012PRL}. We will not analyze this latter method here, but our general conclusions apply to this situation as well.

We start by discussing signatures of quantum motion in Section \ref{sec:SigQuant}, where we also point out that a negative Wigner distribution is not a necessary requirement for detecting nonclassical features. In Section \ref{sec:Setup}, we present the experimental situation and the measurements that we will be analyzing. Section \ref{sec:ModelOptomech} presents the model we use to describe the optomechanical system. We then derive general expressions for the heterodyne photocurrent using different models for both the electromagnetic field and the detector, which clearly shows how the interpretation of sideband asymmetry differs depending on the model. This is presented in Sections \ref{sec:ClassClass}-\ref{sec:QuantQuant}. In Section \ref{sec:Compare}, the two detector models are shown to be equivalent in standard quantum theory, and this is used to derive a relation between quantum optomechanical correlations and the asymmetry in the oscillator's noise spectrum. We then introduce a detailed model for the mechanical oscillator in Section \ref{sec:MechModel} which is used to derive explicit expressions for the sidebands and their asymmetry in Section \ref{sec:Explicit}. Following that, we compare the quantum result with the result of a fully classical theory in Section \ref{sec:ClassAsymm} and point out under which circumstances they can differ. Finally, we conclude in Section \ref{sec:Conclusion}.

\section{Signatures of quantum motion}
\label{sec:SigQuant}

In this section, we discuss how measurements of the noise spectrum of a mechanical oscillator's position can be used to conclude that it behaves according to quantum theory \cite{Clerk2010RMP}. We also discuss how quantum motion can be detected even if the Wigner distribution is everywhere positive.

\subsection{Mechanical oscillator interacting with environment}

Let us consider a mechanical oscillator with mass $m$ in a harmonic potential with an associated angular frequency $\omega_m$. We assume that its position $X$ and momentum $P$ can be found from the following equations:
\begin{align}
\label{eq:EOMMechOsc}
 \dot{X} & = \frac{P}{m}  \\
 \dot{P} & = - m \omega_m^2 X - \frac{\gamma_m}{2} P + F . \notag
\end{align}
We treat the oscillator's interaction with its environment in a Markovian approximation \cite{Gardiner1985PRA,Clerk2010RMP}, leading to an energy damping rate $\gamma_m$, as well as a fluctuating force $F$ on the oscillator. The Markov approximation is usually good in the case of a high-$Q$ oscillator, i.e., when the damping rate $\gamma_m$ is much smaller than the frequency $\omega_m$. This is the case we consider throughout this article.

In classical physics, the position and momentum of a mechanical oscillator are real numbers with definite values at all points in time. In the presence of the noisy force $F$, it is not possible to calculate these definite values for a particular experimental run. The statistical properties of the oscillator can then be described by a phase space probability distribution which depends on the statistical properties of the force $F$. 

In a quantum theory, Equations \eqref{eq:EOMMechOsc} should be interpreted as operator equations, and the position and momentum operators must satisfy the canonical commutation relation
\begin{equation}
\label{eq:CanCommRel}
[X,P] = i \hbar .
\end{equation}
For convenience, we introduce the dimensionless variables
\begin{equation}
\label{eq:Dimlessxp}
x = \frac{X}{X_\mathrm{zpf}} \ , \ p = \frac{P}{P_\mathrm{zpf}} 
\end{equation}
by defining the constants 
\begin{equation}
\label{eq:ZPFDefs}
X_\mathrm{zpf} = \sqrt{\frac{\hbar}{2m\omega_m}} \ , \ P_\mathrm{zpf} = \sqrt{\frac{\hbar m\omega_m}{2}} .
\end{equation}
We note that $X_\mathrm{zpf}$ is the size of the zero point fluctuations in a quantum theory, meaning that $X_\mathrm{zpf}^2 = \langle 0 | X^2 |0\rangle$ where $|0\rangle$ is the ground state. Similarly, we have $P_\mathrm{zpf}^2 = \langle 0 | P^2 |0\rangle$. In terms of the dimensionless variables, Equation \eqref{eq:CanCommRel} translates into $[x,p] = 2 i$. Even though the definitions \eqref{eq:Dimlessxp} arise naturally in quantum mechanics, we will for convenience also use the dimensionless variables $x$ and $p$ when treating the oscillator classically. In that case, $X_\mathrm{zpf}$ ($P_\mathrm{zpf}$) can simply be thought of as an arbitrary length (momentum). 

In the Markov approximation, the noisy force $F$ can be treated as white noise, i.e., its value at a specific time does not depend on its value at other times. Causality then requires $\langle F(\tau) x(0) \rangle = \langle F(\tau) p(0) \rangle = 0$ when $\tau > 0$, both in the classical and quantum case. Here, $\langle \cdots \rangle$ denotes an ensemble average, i.e., an average over different noise configurations.

\subsection{Asymmetric noise spectrum}

The frequency content of the fluctuations in the mechanical oscillator position can be characterized by the noise spectrum
\begin{equation}
\label{eq:DefMechNoiseSpect}
S_{xx}[\omega] = \int_{-\infty}^\infty d \tau \, e^{i \omega \tau} \langle x(\tau) x(0) \rangle .
\end{equation}
For a classical variable $x$, this noise spectrum must be symmetric in frequency, i.e., $S_{xx}[-\omega] = S_{xx}[\omega]$. This is straightforward to show by changing the time variable $\tau \rightarrow -\tau$ and using time translational symmetry $\langle x(-\tau) x(0) \rangle = \langle x(0) x(\tau) \rangle$. The final step $\langle x(0) x(\tau) \rangle = \langle x(\tau) x(0) \rangle$ is trivial for a classical variable $x$. In quantum mechanics, however, $x$ does not necessarily commute with itself at different times. As a consequence, the spectrum need not be symmetric in frequency.

By using the equations of motion and exploiting the white noise nature of $F$, the noise spectrum defined in Equation \eqref{eq:DefMechNoiseSpect} can in the limit $\omega_m/\gamma_m \gg 1$ be expressed as  
\begin{align}
\label{eq:SpectSingleTime}
& S_{xx}[\omega] \\
& =  \frac{|\chi_m[\omega]|^2}{2} \left[\gamma_m\left(\langle x^2 \rangle - \frac{i}{2} \langle [x, p] \rangle \right) - (\omega - \omega_m) \langle \{ x, p \} \rangle  \right] \notag \\
&  \frac{|\chi_m[-\omega]|^2}{2} \left[\gamma_m\left(\langle x^2 \rangle + \frac{i}{2} \langle [x, p] \rangle \right) + (\omega + \omega_m) \langle \{ x, p \} \rangle  \right] \notag .
\end{align}
This is shown in Appendix \ref{app:EqTime}. Here, $\{ \cdot , \cdot \}$ is the anticommutator and we have defined the mechanical susceptibility
\begin{equation}
\label{eq:chimDef}
\chi_m[\omega] = \frac{1}{\gamma_m/2 - i (\omega - \omega_m)} .
\end{equation}
We note that
\begin{equation}
\label{eq:Lorents}
|\chi_m[\pm \omega]|^2 = \frac{1}{(\gamma_m/2)^2 + (\omega \mp \omega_m)^2} 
\end{equation}
is a Lorentzian of width $\gamma_m$ centered at the frequency $\pm \omega_m$.

In a thermal state at temperature $T$, we have 
\begin{equation}
\label{eq:x2Therm}
\langle x^2 \rangle = 2 n_\therm + 1 ,
\end{equation}
where 
\begin{equation}
\label{eq:nTherm}
n_\therm = \frac{1}{e^{\hbar \omega_m/(k_B T)} - 1}
\end{equation}
is the average number of vibrational quanta, i.e., phonons. In the high temperature limit, we get $n_\therm \approx k_B T/(\hbar \omega_m) \gg 1$ and $\langle X^2 \rangle \approx k_B T/(m \omega_m^2)$ in accordance with the classical equipartition principle. Also, for a thermal state, we have $\langle \{x,p\} \rangle = 0$. This gives the thermal noise spectrum
\begin{align}
\label{eq:SpectThermal}
S_{xx}[\omega] & = \gamma_m \left[ (n_\therm + 1)|\chi_m[\omega]|^2 + n_\therm |\chi_m[-\omega]|^2 \right]   .
\end{align}
We see that there is an asymmetry between positive and negative frequencies which is resolvable in the low temperature regime $n_\therm \lesssim 1$. The asymmetry can be traced back to the nonzero commutator in Equation \eqref{eq:CanCommRel} or, equivalently, to the fact that $x$ does not commute with itself at different times. 

\subsection{Quantum motion and the Wigner distribution}

The Wigner distribution $W(x,p)$ is one example of a quasiprobability distribution that can be used to calculate expectation values of quantum operators as phase space integrals over real numbers $x$ and $p$, similarly to how expectation values are calculated in classical statistical mechanics. The Wigner distribution for a single oscillator is defined as 
\begin{equation}
\label{eq:WigDef}
W(x,p) = \frac{1}{2\pi} \int_{-\infty}^\infty dx' \, e^{ipx'} \left\langle x - \frac{x'}{2} \Big| \rho \Big| x + \frac{x'}{2} \right\rangle 
\end{equation}
where $\rho$ is the density matrix. This can be used to calculate expectation values that are symmetric with respect to the order of $x$ and $p$ operators. For example, we have
\begin{equation}
\label{eq:ExpWig1}
\langle x^2 \rangle = \int_{-\infty}^\infty dx \int_{-\infty}^\infty dp \, x^2 W(x,p) 
\end{equation}
as well as 
\begin{equation}
\label{eq:ExpWig2}
\frac{1}{2}\langle \{ x,p \} \rangle = \int_{-\infty}^\infty dx \int_{-\infty}^\infty dp \, x p \, W(x,p) .
\end{equation}
Superficially, this gives the impression that quantum mechanics is nothing more than classical statistical mechanics. However, it is well known that some states have Wigner distributions that are negative in certain regions of phase space, which means that it cannot be interpreted as a probability distribution. 

The fact that negativity hinders an interpretation of the Wigner distribution as a probability distribution seems to have led to a widespread belief that it is only states $\rho$ with negative Wigner distributions that are ''true quantum states''. The idea is that states with everywhere positive Wigner distributions can be described by classical statistical mechanics. This picture is too simplistic, however. In fact, negativity of the Wigner distribution is neither a necessary nor a sufficient condition for the failure of a classical theory \cite{Spekkens2008PRL}. 

The inadequacy of using negativity of the Wigner distribution as a measure of non-classicality can be illustrated by considering a scenario where non-symmetrized expectation values can be measured. Such expectation values cannot be calculated from a Wigner distribution. Consider for example the commutator $[x,p] = xp - px$ that enters the spectrum in \eqref{eq:SpectSingleTime}. This is clearly not symmetric with respect to the order of $x$ and $p$. In other words, a measurement of the noise spectrum $S_{xx}[\omega]$ effectively measures expectation values that cannot be calculated with the Wigner distribution alone. 

We conclude that, with an appropriate detector, it is possible to detect quantum features even if the state has an associated Wigner distribution that is positive at all points in phase space. This is relevant here, because we will be considering a mechanical oscillator in a thermal state (which becomes the ground state in the limit of zero temperature). This type of state has a Gaussian Wigner distribution that is everywhere positive.

\section{Experimental setup and measurement}
\label{sec:Setup}
\subsection{Setup}

\begin{figure}[htbp]
\begin{center} 
\includegraphics[width=.49\textwidth]{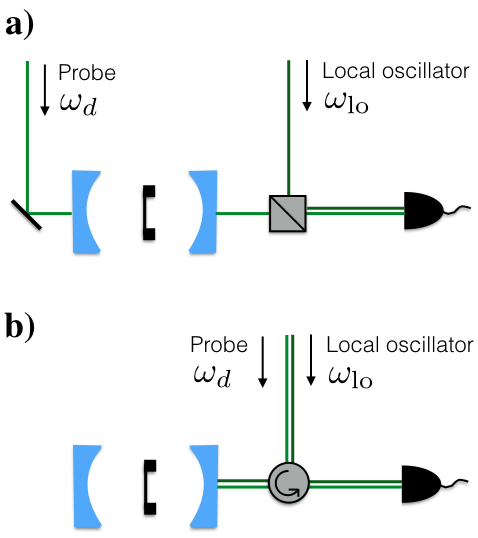}
\caption{The experimental setup that we will have in mind. An optical cavity mode's resonance frequency depends on the motion of a mechanical oscillator, which is here depicted as a thin dielectric membrane. A probe laser beam at frequency $\omega_d$ is sent into the cavity. The light emanating the cavity is combined with a local oscillator beam at frequency $\omega_\mathrm{lo}$. The two beams are detected by a photomultiplier. a) Measurement in transmission where the cavity output and the local oscillator are combined on a beam splitter before arriving at the photomultiplier. b) Measurement in reflection where both beams are sent towards the cavity. The probe beam enters the cavity, whereas the local oscillator is promptly reflected. The beams are then sent to the photomultiplier by means of a circulator.}
\label{fig:Setup}
\end{center}
\end{figure}

We consider an experimental setup as sketched in Figure \ref{fig:Setup}. 
The frequency of an optical cavity mode depends linearly on the position of a mechanical oscillator. The optical mode is driven by a laser beam at (angular) frequency $\omega_d$ for the purpose of measuring the mechanical oscillator position. In a real experiment, an additional laser drive can be present for the purpose of cooling the mechanical oscillator mode. However, since we are only concerned with the detection part here, we can safely ignore the presence of other drives addressing other cavity modes.

Measurements are performed on the light leaking out of the cavity, which contains information about the mechanical oscillator. The theory we present can be applied both to setups where measurements are performed in transmission (as in Figure \ref{fig:Setup}a) and setups where measurements are performed in reflection (as in Figure \ref{fig:Setup}b). We also emphasize that our theory is not limited to a Fabry-P\'{e}rot type cavity as in Figure \ref{fig:Setup}, but is for example also valid for a photonic crystal cavity \cite{Safavi-Naeini2012PRL} or a microtoroidal cavity \cite{Riviere2011PRA}. One should, however, note that if such cavities are driven via evanescent coupling to an optical fiber, measurement in transmission (reflection) is then described by what we here call measurement in reflection (transmission).
 
\subsection{Measuring the heterodyne spectrum} 
 
The light from the cavity is combined with a local oscillator beam at frequency $\omega_\mathrm{lo}$ before detection in a photomultiplier. We imagine that the current $i(t)$ generated in the photomultiplier is recorded for a sampling time $T_s$. This measurement record $i(t)$ is then obviously a classical variable. The windowed Fourier transform \cite{Clerk2010RMP} of the current can be defined as
\begin{equation}
\label{eq:Window}
i_{T_s}[\omega] = \frac{1}{\sqrt{T_s}}\int^{T_s/2}_{-T_s/2} dt \, e^{i \omega t} \, i(t) .
\end{equation}
This can be used to calculate the spectral density of the photocurrent, defined as
\begin{equation}
\label{eq:spectralDens}
S[\omega] = \lim_{T_s \rightarrow \infty} \overline{|i_{T_s}[\omega]|^2} .
\end{equation}
The bar indicates an ensemble average, i.e., an average over many measurements of the absolute square of $i_{T_s}[\omega]$. In practice, one might only use one measurement record $i(t)$ with a large sampling time $T_s$ and assume ergodicity, i.e., that time and ensemble averages are equivalent.

The Wiener-Khinchin theorem states that
\begin{equation}
\label{eq:WKTheorem}
S[\omega] = \int_{-\infty}^{\infty} d \tau \, e^{i\omega \tau} G_{ii}(\tau)
\end{equation}
with the time-averaged autocorrelation function 
\begin{equation}
\label{eq:GiiDef}
G_{ii}(\tau) = \lim_{T_s \rightarrow \infty} \frac{1}{T_s} \int_{-T_s/2}^{T_s/2} dt  \, \overline{i(t + \tau) i(t)} .
\end{equation}
The integrand in Equation \eqref{eq:GiiDef} is an ensemble average over the product of the classical measurement record $i(t)$ at two different times. The question is now how this average relates to expectation values of the optical and mechanical degrees of freedom. The answer depends on which model is used for the photodetector, which again will lead to different interpretations of the features in the photocurrent spectrum. We note that since $G_{ii}(\tau)$ must be symmetric in time $\tau$, we can write
\begin{equation}
\label{eq:WKTheorem2}
S[\omega] = \int_{-\infty}^{\infty} d \tau \, \cos(\omega \tau) G_{ii}(\tau) ,
\end{equation}
which shows that the spectrum $S[\omega]$ is symmetric in frequency.

\begin{figure}[htbp]
\begin{center} 
\includegraphics[width=.5\textwidth]{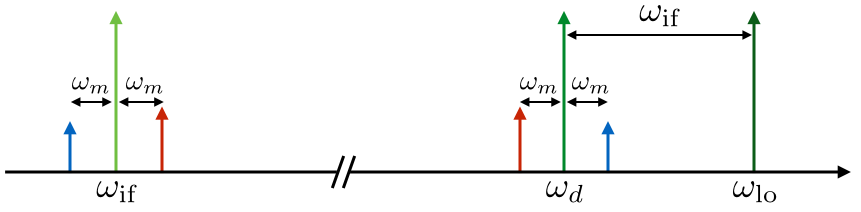}
\caption{An overview of the angular frequencies involved. The probe laser beam at frequency $\omega_d$ will acquire a red-shifted (blue-shifted) sideband at $\omega_d - \omega_m$ $(\omega_d + \omega_m)$ as a result of modulation by the mechanical oscillator. The noise in the probe laser is converted to noise in the photocurrent around the intermediate frequency $\omega_\imf$ due to beating between the probe beam and the local oscillator. Note that the mixing down makes the sidebands switch place when $\omega_\mathrm{lo} > \omega_d$, i.e., the red (blue) sideband is found at $\omega_\imf + \omega_m$ $(\omega_\imf - \omega_m)$.}
\label{fig:Freqs}
\end{center}
\end{figure}

Due to beating between the probe laser beam at $\omega_d$ and the local oscillator beam at $\omega_\mathrm{lo}$, the photocurrent will contain frequencies around the intermediate frequency
\begin{equation}
\label{eq:imfDef}
\omega_\imf = \omega_\mathrm{lo} - \omega_d .
\end{equation}
We will assume $\omega_\mathrm{lo} > \omega_d$, without loss of generality. This means that the sidebands imprinted on the probe beam by the mechanical oscillator will be converted to sidebands at $\omega_\imf \pm \omega_m$ in the photocurrent spectral density $S[\omega]$. See Figure \ref{fig:Freqs} for an overview of the frequencies involved.
We will later calculate the exact form of the motional sidebands in $S[\omega]$. For now, it is sufficient to note that since the sidebands originate from the oscillator motion, their width must be set by the mechanical oscillator linewidth $\gamma_m$ (assuming that the laser linewidth is negligibly small compared to $\gamma_m$). 

Since we assume $\omega_\mathrm{lo} > \omega_d$, the sideband due to red-shifted light at $\omega_d - \omega_m$ will be mixed down to the frequency $\omega_\mathrm{if} + \omega_m$, whereas the blue sideband at $\omega_d + \omega_m$ will be mixed down to $\omega_\mathrm{if} - \omega_m$. In other words, the sidebands change place when mixing down from optical to RF frequencies. We therefore define the spectrum at the red sideband as 
\begin{equation}
\label{eq:SrrDef}
S_{rr}[\tilde{\omega}] = S[\omega_\mathrm{if} + \tilde{\omega}] 
\end{equation}
with $\tilde{\omega} \sim \omega_m$. Similarly, we will refer to\begin{equation}
\label{eq:SbbDef}
S_{bb}[\tilde{\omega}] = S[\omega_\mathrm{if} - \tilde{\omega}] 
\end{equation}
as the spectrum at the blue sideband.

\section{Model of the optomechanical system}
\label{sec:ModelOptomech}
\subsection{System Hamiltonian}

We now introduce the model for the optomechanical system. To simplify the formalism, we will neglect polarization throughout this article. Since polarization plays no role in the system we study, neglecting it cannot change the result in any significant way. We will thus only be concerned with one mode of the cavity field and will ignore all the other cavity modes. In a classical theory, we denote the cavity mode amplitude by the complex number $a$. In a quantum theory, $a$ will be the photon annihilation operator. See Appendix \ref{app:Formalism} for details regarding the formalism. The cavity mode's angular frequency will be denoted $\omega_{c}$. Note that we will, for simplicity, define $\omega_c$ such that it includes a slight shift due to the average displacement experienced by the mechanical oscillator when the laser is on.

We assume that the cavity mode frequency depends linearly on the position $X$ of a mechanical oscillator mode for a sufficiently large range of positions. We also let the cavity mode be driven by a laser at the drive frequency $\omega_d$. This leads to the system Hamiltonian
\begin{align}
\label{eq:HsDef}
H_\mathrm{sys} & = \hbar \left[\omega_{c} + A (X - X_0) \right] a^\ast a + H_\mathrm{mech} \notag \\ & + i \hbar \Omega \left[ e^{-i \omega_d t} (1 + n) a^\ast - e^{i \omega_d t} (1 + n^\ast) a \right]
\end{align}
where $\omega_{c}$ is the resonance frequency of the cavity mode when $X = X_0$ and $X_0$ is the equilibrium position of the mechanical oscillator when the laser beam is on. We have defined the parameter $A = (\partial \omega_c / \partial X)|_{X = X_0}$. The Hamiltonian for the mechanical mode $H_\mathrm{mech}$ will be discussed in Section \ref{sec:MechModel}, but can be left unspecified for now. The real parameter $\Omega$ is proportional to $\sqrt{P}$, where $P$ is the power of the laser drive. We have introduced a complex and dimensionless variable $n(t)$ with $|n(t)| \ll 1$ to describe laser noise, and its properties will be specified later. 

Note that the presence of $\hbar$ in Eq.~\eqref{eq:HsDef} does not imply that the cavity field or the mechanical oscillator mode have been quantized. We have written down the classical Hamiltonian, meaning that $a$ is a complex number. However, the same Hamiltonian can be used for a quantized field with $a^\ast \rightarrow a^\dagger$. See Appendix \ref{app:Formalism} for further details.

\subsection{Coupling to external modes}

The field outside the cavity can be written as an expansion in terms of mode functions. Let us for example think of the electromagnetic field as being linearly polarized, and let $E(\mathbf{r},t)$ be the electric field's scalar value along a particular direction. We write this as
\begin{equation}
\label{eq:EPlusMinus}
E(\mathbf{r},t) = E^{(+)}(\mathbf{r},t) + E^{(-)}(\mathbf{r},t)
\end{equation}
where
\begin{align}
\label{eq:EoutputPlus2}
 E^{(+)}(\mathbf{r},t)  &  
= \sum_j \sqrt{\frac{\hbar \omega_j}{2\varepsilon_0}} b_j(t) w_j(\mathbf{r})  
\end{align}
and $E^{(-)}(\mathbf{r},t) = (E^{(+)}(\mathbf{r},t))^\ast$ (for a classical field). We will refer to $w_j(\mathbf{r})$ as bath mode functions where the mode index $j$ can represent a set of quantum numbers. The bath modes are not necessarily traveling waves, since an electromagnetic field mode terminating at a cavity mirror would form a standing wave along the cavity axis. We will not be too concerned with the details of the bath mode functions, but we note that they are of importance when relating the electromagnetic field in the cavity to the field at the photodetector (see Appendix \ref{app:Edet}).

We can again express the Hamiltonian as a collection of harmonic oscillators with associated coefficients $b_j$ and frequencies $\omega_j$ that now form a continuum. The Hamiltonian of the field is
\begin{equation}
\label{eq:ElmagCont}
H_\mathrm{free} = \sum_j \hbar \omega_j b^\ast_j b_j  .
\end{equation} 
In a quantum theory, $b_j, b_j^\ast \rightarrow b_j, b_j^\dagger$ are operators that satisfy standard bosonic commutation relations.

We will assume that the cavity mode couples to the electromagnetic field modes outside the cavity in the standard bilinear way. For coupling through a single mirror, we would then add 
\begin{align}
\label{eq:Hbath}
H_\mathrm{ext} & = \hbar \sum_j \mu_j \left(a^\ast b_{j} + b_{j}^\ast a \right) + H_\mathrm{free}
\end{align}
to the system Hamiltonian \eqref{eq:HsDef}, where $\mu_j$ is the coupling rate to the bath mode $b_j$. 

In general, there are more than one channel through which the cavity mode can decay. Examples of other decay channels could be the partial transparency of the other mirror, scattering out of the cavity mode, or photon absorption in the mirrors. Similar terms as in Eq.~\eqref{eq:Hbath} could then be added for every decay channel, involving other, independent sets of bath modes.

%

\subsection{Adiabatic elimination of the external modes}

In a standard Markovian treatment of the coupling to the external baths \cite{Gardiner1985PRA,Clerk2010RMP}, one assumes that the bath density of states $D$ and the coupling $\mu_j \rightarrow \mu$ are approximately constant in a sufficiently wide range of bath mode frequencies around the cavity mode frequency. One then ends up with a Langevin equation for the cavity mode coefficient $a$:
\begin{align}
\label{eq:aLangevin}
\dot{a} & = -\left(\frac{\kappa}{2} + i \omega_{c}  \right) a - i A (X - X_0) a \\
& + \left[\Omega + r \sqrt{\kappa} \, \zeta + \sqrt{\kappa_\mathrm{ext}} \, \xi_\mathrm{ext} + \sqrt{\kappa_\mathrm{int}} \, \xi_\mathrm{int} \right] e^{-i \omega_d t} . \notag
\end{align}
The adiabatic elimination of bath variables have introduced the parameter $\kappa$, which is the rate at which energy decays from the cavity mode. Additionally, the coupling to the bath modes gives rise to noise. We have separated the noise into three terms. The first is the laser noise, which has been expressed by the variable
\begin{equation}
\label{eq:ZetaDef}
\zeta(t) = \frac{\Omega}{r\sqrt{\kappa}}  n(t),
\end{equation}
where $r = \Omega/\Omega_0$ is dimensionless and $\Omega_0$ corresponds to some reference value for the laser power $P$. The reason for introducing $r$ is that it scales with $\sqrt{P}$, such that we can expect the fluctuating variable $\zeta(t)$ to be independent of laser power. 

The variable $\xi_\mathrm{ext}(t)$ is the input noise from the bath modes on which we will perform measurements. This represents intrinsic noise in the electromagnetic field that would be there even in absence of the laser drives. We can relate it to the bath modes by
\begin{equation}
\label{eq:xiDef}
\xi_\ext(t) = - \frac{i}{\sqrt{2 \pi D}}  \sum_j e^{-i (\omega_j - \omega_d) (t-t_0)} b_{j}(t_0),
\end{equation}
where $t_0$ is an arbitrary time in the distant past at which we can assume that the cavity and bath modes were uncorrelated. The input noise \eqref{eq:xiDef} is a sum over the coefficients $b_j$ propagated freely from $t_0$ to the time $t$ and represents the noise impinging on the cavity mode from these bath modes \cite{Gardiner1985PRA,Clerk2010RMP}. We note that the sum in Eq.~\eqref{eq:xiDef} is in reality limited to modes $b_j$ for which the coupling rate $\mu_j$ is appreciable. The variable $\xi_\mathrm{int}(t)$ is the noise associated with all other dissipation channels, and can be defined in a similar manner. We will specify the correlation properties of the input noise variables $\zeta$, $\xi_\ext$, and $\xi_\innt$ later. 

The parameter $\kappa_\mathrm{ext} = 2 \pi D  \mu^2$ is the rate at which the cavity mode energy leaks into the bath modes $b_j$, whereas $\kappa_\mathrm{int}$ is the rate for leaking into all other loss channels. We have the relation
\begin{equation}
\label{eq:kappadef}
\kappa = \kappa_\ext + \kappa_\innt 
\end{equation}
for the total energy decay rate.

\subsection{Linearization and calculation of cavity mode fluctuations}

We now move to a frame rotating at the drive frequency $\omega_d$, i.e.~we let $a = e^{-i \omega_d t} \tilde{a}$ and rename $\tilde{a} \rightarrow a$. This gives
\begin{align}
\label{eq:aLangevin2}
\dot{a} & = -\left(\frac{\kappa}{2} - i \Delta \right) a - i g_0 x a \\
& + \Omega + r \sqrt{\kappa} \, \zeta + \sqrt{\kappa_\mathrm{ext}} \, \xi_\mathrm{ext} + \sqrt{\kappa_\mathrm{int}} \, \xi_\mathrm{int}  . \notag
\end{align}
where $\Delta = \omega_d - \omega_{c}$ is the detuning between the laser and the cavity resonance frequency. The detuning will be kept general for now, but we will later focus on the resonant case $\Delta = 0$ (or in practice, $|\Delta| \ll \kappa$). We have also introduced the dimensionless position fluctuation
\begin{equation}
\label{eq:xDimless}
x = \frac{X - X_0}{X_\mathrm{zpf}} .
\end{equation} 
The length $X_\mathrm{zpf}$ can be chosen arbitrarily, but it is convenient to choose $X_\mathrm{zpf} = \sqrt{\hbar/(2 m \omega_m)}$, which is the size of the zero point fluctuations of a quantum harmonic oscillator. The optomechanical coupling rate is then
\begin{equation}
\label{eq:g0Def}
g_0 = A X_\mathrm{zpf} .
\end{equation} 
In quantum cavity optomechanics, this is what is called the single-photon coupling rate. Note, however, that our choice of $X_\mathrm{zpf}$ does not imply that we are doing quantum mechanics. The equation \eqref{eq:aLangevin2} can still be considered a classical Langevin equation.

Next, we write the variable $a$ as a sum of its constant expectation value $\bar{a} = \langle a(t) \rangle$ and a fluctuating part $d(t)$, such that  
\begin{equation}
\label{eq:dDef}
a(t) = \bar{a} + d(t) .
\end{equation} 
We will consider the experimentally relevant limit where the changes in cavity frequency caused by the motion of the oscillator is small compared to the cavity linewidth $\kappa$. In this case, the expectation value $\bar{a}$ is to a good approximation given by
\begin{equation}
\label{eq:abarDef}
\bar{a} = \frac{\Omega}{\kappa/2 - i \Delta}.
\end{equation}
In the same limit, we may also linearize the equation of motion for the cavity mode fluctuations $d(t)$:
\begin{align}
\label{eq:dLangevin}
\dot{d} & = -\left(\frac{\kappa}{2} - i \Delta \right) d - i G x \\
& + r \sqrt{\kappa} \, \zeta + \sqrt{\kappa_\mathrm{ext}} \, \xi_\mathrm{ext} + \sqrt{\kappa_\mathrm{int}} \, \xi_\mathrm{int}  . \notag
\end{align}
Here, we have defined the enhanced (or many-photon) optomechanical coupling
\begin{equation}
\label{eq:GDef}
G = g_0 \bar{a}
\end{equation}
and neglected the term proportional to $x d$.

We define the Fourier transform as
\begin{equation}
\label{eq:Fourier}
h[\omega] = \int_{-\infty}^\infty d t \, e^{i \omega t} h(t)
\end{equation}
for an arbitrary function $h$. For convenience, we define the Fourier transform of the complex conjugate of $h$ as
\begin{equation}
\label{eq:Fourier2}
h^\dagger[\omega] = \int_{-\infty}^\infty d t \, e^{i \omega t} h^\ast(t)
\end{equation}
We note that $h^\dagger[\omega] = (h[-\omega])^\ast$ and emphasize that the dagger superscript is merely a convenient notation at this point and does not necessarily indicate that we have introduced quantum physics. 

In the Fourier domain, the cavity mode fluctuation $d$ becomes
\begin{equation}
\label{eq:dFourier}
d[\omega] = \chi_c[\omega] \left(d_\mathrm{in}[\omega] - i G x[\omega] \right)
\end{equation}
having defined the input cavity noise as
\begin{equation}
\label{eq:dinDef}
d_\mathrm{in}[\omega] = r \sqrt{\kappa} \, \zeta[\omega] + \sqrt{\kappa_\ext} \xi_\ext[\omega] + \sqrt{\kappa_\innt} \xi_\innt[\omega] 
\end{equation}
and the cavity susceptibility as
\begin{equation}
\label{eq:chicDef}
\chi_c[\omega] = \frac{1}{\kappa/2 - i (\omega + \Delta)} . 
\end{equation}
In other words, the variable $d[\omega]$ tells us the cavity mode fluctuations given the known input noise from the reservoirs as well as the noise in the mechanical oscillator position.

\subsection{Commutation relations}
\label{sec:CommReld}

In a fully classical theory, $d$ is a complex number, whereas in standard quantum theory it obeys the bosonic commutation relation $[d,d^\dagger] = 1$. However, we will in the following not specify which commutation relation $d$ must satisfy. We will rather view Equation \eqref{eq:dFourier} as our starting point and specify the properties of the noise variables entering $d_\mathrm{in}$. This means that the commutation relations involving $d$ depends on the properties of the mechanical oscillator, which we do not wish to assume anything about.

In a theory where we (for example) let the input noise $d_\mathrm{in}$ be classical and where $x$ is quantum, this can lead to strange properties in which e.g.~$[d,x] \neq 0$. However, we will not concern ourselves with such aspects. Our focus will simply be to try to model the experiment in the most classical way possible without any assumptions on the nature of the mechanical oscillator.

This means in other words that even if we choose the electromagnetic input field to be classical, the operator $d$ may satisfy nontrivial commutation. From now on, we will therefore treat it as a quantity which we are not allowed to commute with any other variables (and write $d^\ast \rightarrow d^\dagger$) .

\subsection{The output mode}

It is the electric field at the photodetector outside the cavity that will be subject to measurements. We are interested in relating the measurement record to the intracavity field, and we will see that this can be done in terms of what is called the cavity output mode. We define it as \cite{Gardiner1985PRA,Clerk2010RMP} 
\begin{equation}
\label{eq:aOutput}
a_\mathrm{out}(t) = \frac{i}{\sqrt{2 \pi D}}  \sum_j e^{-i (\omega_j-\omega_d) (t-t_1)} b_{j}(t_1),
\end{equation}
where $t_1$ is a time in the distant future, and the sum is again limited to bath modes $b_j$ that couple significantly to the cavity mode.

The output mode can be written as
\begin{equation}
\label{eq:aOutput2}
a_\mathrm{out}(t) = \bar{a}_\mathrm{out} + d_\mathrm{out}(t) ,
\end{equation}
where the constant term  $\bar{a}_\mathrm{out} = \langle a_\mathrm{out}(t) \rangle $ quantifies the strength of the laser beam emanating the cavity from this port and is given by
\begin{equation}
\label{eq:aoutBar}
\bar{a}_\mathrm{out} = \sqrt{\kappa_\ext} \left(\bar{a} - \lambda \frac{\Omega}{\kappa_\ext} \right) .
\end{equation}
Here, we introduce the parameter $\lambda$ that depends on details of the experimental setup. For measurement in transmission, as depicted in Figure \ref{fig:Setup}a, i.e., if the cavity is driven through a different port than the one used for measurement, we have $\lambda = 0$. For measurement in reflection (Figure \ref{fig:Setup}b), i.e., if the cavity is driven and measured through the same port, we have $\lambda = 1$. 

The value of $\bar{a}_\mathrm{out}$ is however not of importance to us. We are interested in the fluctuating part $d_\mathrm{out}$ of the output mode, which in the Fourier domain becomes 
\begin{equation}
\label{eq:dOutput}
d_\mathrm{out}[\omega] = \Gamma[\omega] - i \sqrt{\kappa_\ext} \, G \, \chi_c[\omega] \, x[\omega] .
\end{equation}
Here, all the cavity noise terms that do not contain the oscillator position $x$ have been lumped into
\begin{equation}
\label{eq:Gamma}
\Gamma[\omega] = \sqrt{\kappa_\ext} \chi_c[\omega] d_\mathrm{in}[\omega] - \xi_\ext[\omega] - \lambda r \sqrt{\frac{\kappa}{\kappa_\ext}} \zeta[\omega] .
\end{equation}
We have now established how the cavity output depends on the mechanical oscillator and the input noise from the electromagnetic field. Note that  the theory, as well as the results we present in Sections \ref{sec:ClassClass}-\ref{sec:QuantQuant}, is valid for any degree of freedom $x$ that couples linearly to electromagnetic field fluctuations, not just a mechanical oscillator. 

Finally, let us point out that for the same reasons discussed in Section \ref{sec:CommReld}, we cannot make any assumptions about which commutation relations the output mode $a_\mathrm{out}$ should satisfy.

\subsection{The detector mode}

Let us for a moment consider the setup in Figure \ref{fig:Setup}a, where the output from the cavity is combined with the local oscillator on a beam splitter. The electric field at the photodetector is then a linear combination of the field propagated from the cavity and the field from the local oscillator. If we wish to denote the mode coefficients of the field at the detector by $b_j$, Equation \eqref{eq:aOutput} should be redefined by replacing $a_\mathrm{out}$ with $a_\mathrm{det}$, where we define the detector mode as 
\begin{equation}
\label{eq:adet}
a_\mathrm{det}(t) = \sqrt{T} a_\mathrm{out}(t) - i \sqrt{1-T} e^{-i \omega_\mathrm{if} t} a_\mathrm{lo}(t)  ,
\end{equation} 
the local oscillator mode as
\begin{equation}
\label{eq:alo}
a_\mathrm{lo}(t) = \bar{a}_\mathrm{lo} + d_\mathrm{lo}(t)  ,
\end{equation}
and the output mode $a_\mathrm{out}$ according to Equation \eqref{eq:aOutput2}. The constant $T$ is the intensity transmission coefficient of the beam splitter. The constant $\bar{a}_\mathrm{lo}$ describes the local oscillator beam and $d_\mathrm{lo}(t)$ represents noise. We let $\bar{a}_\mathrm{lo}$ be real, without loss of generality. 

We will also assume that Equation \eqref{eq:adet} is valid for the setup in Figure \ref{fig:Setup}b. There is no beam splitter in that setup, but we could transform our model to describe that setup by simply redefining $-i\sqrt{1-T} a_\mathrm{lo} \rightarrow a_\mathrm{lo}$ and $\sqrt{T} a_\mathrm{out} \rightarrow a_\mathrm{out}$. Since this does not lead to any significant differences from the case with a beam splitter, we will work with Equation \eqref{eq:adet} in the following.

\section{Classical field and classical detector}
\label{sec:ClassClass}

Our primary goal is to derive general expressions for the photocurrent spectral density $S[\omega]$. To do this, we will need to specify the properties of the electromagnetic field input noise. In addition, we need to choose a detector model, i.e., we need to relate the photocurrent autocorrelation function $G_{ii}(\tau)$ in Equation \eqref{eq:GiiDef} to expectation values of optical and mechanical degrees of freedom. We will first consider a classical model both for the electromagnetic input field and for the photodetector.

\subsection{Electromagnetic field noise}

We now need to define the properties of the noise variables $\zeta$, $\xi_\ext$, and $\xi_\innt$. The laser noise $\zeta$ will always be considered classical, i.e., a complex number, and we express this as
\begin{equation}
\label{eq:dxdyDef}
\zeta(t) = \frac{1}{2} \left(\delta x(t) + i \delta y(t) \right) .
\end{equation}
The relation between the dimensionless laser noise $n(t)$ and the dimensionful {\it amplitude} noise $\delta x (t)$ and {\it phase} noise $\delta y (t)$ is determined by Equation \eqref{eq:ZetaDef}. The amplitude and phase noise variables are assumed to obey the properties
\begin{align}
\label{eq:classnoiseDef}
\langle \delta x[\omega] \delta x[\omega'] \rangle & = 2 \pi C_{xx}[\omega] \delta(\omega + \omega') \notag \\
\langle \delta x[\omega] \delta y[\omega'] \rangle & = 2 \pi C_{xy}[\omega] \delta(\omega + \omega') \\
\langle \delta y[\omega] \delta y[\omega'] \rangle & = 2 \pi C_{yy}[\omega] \delta(\omega + \omega') \notag
\end{align}
where the functions $C_{xx}, C_{yy}$ must be symmetric in frequency and the Cauchy-Schwarz inequality $|C_{xy}[\omega]|^2 \leq C_{xx}[\omega] C_{yy}[\omega]$ must be satisfied. We will assume that the functions $C_{ij}[\omega] \approx C_{ij}$ are approximately constant over a region of width $\gamma_m$ around the mechanical frequency $\omega_m$, where $\gamma_m$ is the linewidth of the mechanical oscillator. In other words, the laser noise is modeled as white and Gaussian noise in the frequency range of interest. For convenience, we note that
\begin{align}
\label{eq:classnoiseDef2}
\langle \zeta^\dagger[\omega] \zeta[\omega'] \rangle & = \frac{\pi}{2}\left(C_{xx}[\omega] + C_{yy}[\omega]\right) \delta(\omega + \omega') \\
\langle \zeta[\omega] \zeta[\omega'] \rangle & = \frac{\pi}{2}\left(C_{xx}[\omega] + 2i C_{xy}[\omega] - C_{yy}[\omega]\right) \delta(\omega + \omega') . \notag
\end{align}

In this section, we also consider the intrinsic noise variables $\xi_\ext$ and $\xi_\innt$ to be classical. It is natural to assume a similar but more symmetric model where amplitude and phase noise are uncorrelated and where the noise in the phase and amplitude quadratures have the same size. This means that
\begin{align}
\label{eq:classnoiseDef3}
\langle \xi_i^\dagger[\omega] \xi_j[\omega'] \rangle & =  \pi \alpha[\omega] \, \delta_{i,j} \delta(\omega + \omega') \\
\langle \xi_i[\omega] \xi_j[\omega'] \rangle & =  0 , \notag
\end{align}
where the indices $i,j$ represent either {\it ext} or {\it int}. We assume that the intrinsic noise is sufficiently broadband such that for frequencies $\omega \ll \omega_c$, we have $\alpha[\omega] \approx \alpha$, where $\alpha$ is a real and positive constant. 

We should also specify the properties of the noise $d_\mathrm{lo}(t)$ in the local oscillator. In general, we can write this as a combination of laser noise $\zeta_\mathrm{lo}(t)$ and intrinsic field noise $\xi_\mathrm{lo}(t)$, i.e.
\begin{equation}
\label{eq:LoNoise}
d_\mathrm{lo}(t) = r_\mathrm{lo} \zeta_\mathrm{lo}(t) + \xi_\mathrm{lo}(t) .
\end{equation}
Again, we let the constant $r_\mathrm{lo}$ scale as square root of local oscillator laser power. It is reasonable to assume that the intrinsic noise $\xi_\mathrm{lo}(t)$ is uncorrelated with the noise of the cavity output mode. We can therefore extend the indices $i,j$ in \eqref{eq:classnoiseDef3} to also include {\it lo}. However, it is possible that the laser noise $\zeta_\mathrm{lo}(t)$ could be correlated with the output mode if the two beams originate from the same laser. Since the laser noise scales with square root of power, this must then be taken into account, and Ref.~\cite{Jayich2012NJP} contains a theoretical treatment of this case. However, including local oscillator laser noise will not affect the interpretation of sideband asymmetry that we will discuss here. Furthermore, we will later focus on the experimentally relevant scenario where laser noise can be sufficiently removed by filtering. We therefore neglect local oscillator laser noise in the following.

\subsection{Detector model}

In a classical model for the photodetector, it is natural to assume that the average photocurrent $\overline{i(t)}$ is proportional to the expectation value of the square of the electric field averaged over the size of the detector. In other words, we assume that $\overline{i(t)} = \langle I(t) \rangle$ with the stochastic variable $I$ defined as
\begin{equation}
\label{eq:iClass}
I(t) = \frac{q}{2} \int d \mathbf{r} \, f(\mathbf{r})  E^2(\mathbf{r},t) 
\end{equation}
with some proportionality constant $q$. Here, $\langle \cdots \rangle$ is an ensemble average over different configurations of the noise both in the electromagnetic field and in the mechanical oscillator's intrinsic environment. The dimensionless function $f(\mathbf{r})$ has support only in the active region of the detector. The expression \eqref{eq:iClass} assumes that the detector bandwidth is much larger than the relevant frequencies contained in the photocurrent. We also assume that the gain of the detector is frequency independent in the relevant range of frequencies (or that one can accurately compensate for any differences between the gain at the upper and lower sidebands).

We would now like to express the variable $I(t)$ in terms of the detector mode $a_\mathrm{det}$. In Appendix \ref{app:Edet}, we show that 
\begin{equation}
\label{eq:iadet}
I(t) =  \frac{q v^2}{2} \left(a_\mathrm{det}(t) a^\dagger_\mathrm{det}(t) + a^\dagger_\mathrm{det}(t) a_\mathrm{det}(t) \right) 
\end{equation}
to a good approximation, where $v$ is a constant. This is based on the fact that we are only concerned with modes of the electric field with frequencies in a narrow range of width $\sim \omega_\mathrm{if} \ll \omega_d$. We have also neglected terms oscillating at $\pm 2 \omega_d$ since the detector cannot react at such high frequencies. For a classical variable $a_\mathrm{det}$, for which $a_\mathrm{det}^\dagger = a_\mathrm{det}^\ast$, the two terms in Equation \eqref{eq:iadet} are of course the same. However, even if we assume that the input noise of the electromagnetic field is classical for now, we will not assume {\it a priori} that the detector mode $a_\mathrm{det}$ is a classical variable. The reason is that we do not want to make any assumptions on whether the mechanical oscillator is classical or not (see Section \ref{sec:CommReld}). 

In the following, we consider the limits
\begin{align}
\label{eq:AssumptionsBS}
1 - T & \ll 1  \\
|\bar{a}_\mathrm{out}|^2 T & \ll \bar{a}_\mathrm{lo}^2 (1 - T) . \notag
\end{align}
The first assumption is simply that the beam splitter intensity transmission coefficient is close to 1. The second assumption means that the power of the part of the local oscillator beam that arrives at the detector is much larger than the power of the output beam from the cavity. This means that the contribution from the intrinsic noise $\xi_\mathrm{lo}$ of the local oscillator can be disregarded. With these assumptions, we can make the approximation
\begin{equation}
\label{eq:iadet2}
I(t) =  i_0 + Z \left(e^{i \omega_\mathrm{if} t } a_\mathrm{out}(t) - e^{-i \omega_\mathrm{if} t } a^\ast_\mathrm{out}(t) \right)
\end{equation}
where 
\begin{equation}
\label{eq:i0Def}
i_0 = qv^2 (1-T) \bar{a}_\mathrm{lo}^2
\end{equation}
is a constant contribution to the photocurrent and we have defined
\begin{equation}
\label{eq:ZDef}
Z = i q v^2 \sqrt{T(1-T)} \bar{a}_\mathrm{lo} .
\end{equation}

In a classical model for the photodetector, any noise in the photocurrent can only come from the noise in the electromagnetic field at the detector, not from the detection process itself (ignoring, of course, electrical noise originating elsewhere in the measurement apparatus). This means that the ensemble average that enters the photocurrent autocorrelation function in Equation \eqref{eq:GiiDef} simply becomes 
\begin{equation}
\label{eq:iiCorrClass}
\overline{i(t+\tau) i(t)} = \langle I(t + \tau) I(t) \rangle
\end{equation}
where, again, $\langle \cdots \rangle$ is an ensemble average over different configurations of the noise in the electromagnetic field and in the mechanical oscillator's intrinsic environment.

\subsection{Heterodyne spectrum and sideband asymmetry}

We will now express the photocurrent noise spectrum $S[\omega]$ defined  in Equation \eqref{eq:WKTheorem2} in terms of the electromagnetic input noise and the noise in the mechanical oscillator position. In doing this, we will make repeated use of the notation
\begin{equation}
\label{eq:SAB}
S_{AB}[\omega] = \int_{-\infty}^\infty d\tau \, e^{i \omega \tau} \langle A(\tau) B(0) \rangle 
\end{equation}
for arbitrary variables $A$ and $B$. 

We find that for frequencies $\omega = \omega_\imf + \tilde{\omega}$ with $\tilde{\omega} \sim \pm \omega_m$, i.e., in the vicinity of the sideband frequencies, we can write the photocurrent noise spectrum as
\begin{equation}
\label{eq:Sclassclass}
S[\omega] = S^{(o)}[\omega] + S^{(om)}[\omega] + S^{(m)}[\omega] .
\end{equation}
The first term is a noise floor resulting from broadband noise in the electromagnetic field and is given by 
\begin{align}
\label{eq:Sclassclass-o}
S^{(o)}[\omega] & = |Z|^2 \left[ \alpha  + \frac{r^2 \kappa}{4 \kappa_\ext} |\kappa_\ext \chi_c[-\tilde{\omega}] - \lambda|^2 \left(C_{xx} + C_{yy} \right) \right] .  
\end{align}
The first term in the brackets comes from the intrinsic noise of the field and the second from laser noise. We have assumed that the laser noise coefficients $C_{ij}[\tilde{\omega}]$ can be approximated by a constant $C_{ij}$, and that $C_{ij}[2\omega_\mathrm{if}] \approx 0$. The latter approximation means that, unlike the intrinsic noise, the laser noise does not contribute to the image sideband.  

The second term in Eq.~\eqref{eq:Sclassclass} is given by
\begin{align}
\label{eq:Sclassclass-om}
S^{(om)}[\omega] & = -2 |Z|^2 \sqrt{\kappa_\ext} \, \mathrm{Im} \big(G^\ast \chi_c^\ast[-\tilde{\omega}] S_{x\Gamma}[\tilde{\omega}] \big) .
\end{align}
The spectrum $S_{x\Gamma}[\tilde{\omega}]$ is defined according to Equation \eqref{eq:SAB}. We see that the term $S^{(om)}[\omega]$ originates from the optomechanical correlation between the classical electromagnetic field noise $\Gamma$, as defined in Eq.~\eqref{eq:Gamma}, and the position $x$ of the mechanical oscillator.

The last term in Eq.~\eqref{eq:Sclassclass} reads
\begin{align}
\label{eq:Sclassclass-m}
S^{(m)}[\omega] & = |Z|^2 \kappa_\ext |G|^2 |\chi_c[-\tilde{\omega}]|^2 \bar{S}_{xx}[\tilde{\omega}]  
\end{align}
when expressed in terms of the {\it symmetrized} noise spectrum of the mechanical oscillator position, defined by
\begin{equation}
\label{eq:symmSxx}
\bar{S}_{xx}[\tilde{\omega}] = \frac{1}{2} \big( S_{xx}[\tilde{\omega}] + S_{xx}[-\tilde{\omega}] \big) .
\end{equation}
We see that the detector model we have used here leads to a photocurrent spectrum $S[\omega]$ that is {\it not} explicitly susceptible to asymmetry between positive and negative frequencies in the mechanical oscillator noise spectrum $S_{xx}[\tilde{\omega}]$.
  
We now let $\tilde{\omega} \sim \omega_m$ and define the asymmetry function $\Delta S[\tilde{\omega}]$ by
\begin{equation}
\label{eq:DeltaSDef}
\Delta S[\tilde{\omega}] = S[\omega_\imf + \tilde{\omega}] - S[\omega_\imf - \tilde{\omega}] = S_{rr}[\tilde{\omega}] - S_{bb}[\tilde{\omega}].
\end{equation}
This function captures the difference between the red and blue sidebands. Using the expressions above, $\Delta S[\tilde{\omega}]$ becomes
\begin{align}
\label{eq:asymmetryClassClass}
\Delta S[\tilde{\omega}] & = -  |Z|^2 \sqrt{\kappa_\ext}  \\
& \times \big\{ 2 \, \mathrm{Im} \big[G^\ast (\chi_c^\ast[-\omega_m] S_{x\Gamma}[\tilde{\omega}] - \chi_c^\ast[\omega_m] S_{x\Gamma}[-\tilde{\omega}]) \big] \notag \\
&  \quad -  \sqrt{\kappa_\ext} |G|^2 \left(|\chi_c[-\omega_m]|^2 - |\chi_c[\omega_m]|^2 \right) \bar{S}_{xx}[\tilde{\omega}]  \big\}. \notag
\end{align}
We have approximated $\tilde{\omega}$ by $\omega_m$ in the cavity susceptibility, which is fine as long as $\tilde{\omega} - \omega_m \sim \gamma_m \ll \kappa$. In other words, as long as the width of the sidebands, which is set by the mechanical linewidth $\gamma_m$, is much smaller than the cavity linewidth $\kappa$, this is a good approximation. 

The last line in Equation \eqref{eq:asymmetryClassClass} shows that sideband asymmetry can occur if the cavity filters the red and blue sidebands to differing degrees, i.e., if $|\chi_c[-\omega_m]|^2 \neq |\chi_c[\omega_m]|^2$. However, this can be avoided by choosing zero detuning between the laser and the cavity resonance frequency, i.e., $\Delta = 0$, in which case $\chi^\ast_c[-\omega] = \chi_c[\omega]$. In practice, one needs to control $\Delta$ to an accuracy much smaller than the cavity linewidth $\kappa$ in order to rule out that the sideband asymmetry is simply caused by cavity filtering.
 
Let us now assume $\Delta = 0$, which gives 
\begin{align}
\label{eq:asymmetryClassClass2}
\Delta S[\tilde{\omega}] & = - 2 |Z|^2 \sqrt{\kappa_\ext} G \\
& \times  \mathrm{Im} \left( \chi_c[\omega_m] S_{x\Gamma}[\tilde{\omega}] - \chi_c^\ast[\omega_m] S_{x\Gamma}[-\tilde{\omega}]  \right) \notag .
\end{align}
Note that $G$ is real when $\Delta = 0$. We now see that the asymmetry function contains no terms proportional to the oscillator spectrum $S_{xx}[\tilde{\omega}]$. We conclude that at zero laser detuning, the sideband asymmetry in this model would have to originate from the classical optomechanical correlation $S_{x\Gamma}[\tilde{\omega}]$. In Section \ref{sec:ClassAsymm}, we will return to this classical correlation and investigate whether or not it can reproduce the sidebands one would expect in a quantum theory with no classical noise.

\subsection{Failure of a classical model for intrinsic field noise}

It is well known that the classical model \eqref{eq:classnoiseDef3} is not a viable model for the intrinsic noise of the electromagnetic field in the absence of thermal photons, i.e., in the (experimentally relevant) temperature regime where $k_B T \ll \hbar \omega_c$. The reason is that it leads to both nonzero photocurrent and nonzero photocurrent noise even in the case when there are no light sources. If both lasers were turned off, i.e., if $\bar{a}_\mathrm{lo} = \bar{a}_\mathrm{out} = 0$, the classical model we have used so far would give the flat photocurrent noise spectrum  
\begin{equation}
\label{eq:Snolaser}
S[\omega] = B \left(\frac{q v^2 \alpha}{2}  \right)^2
\end{equation}
if we assume that the detector bandwidth $B$ is much smaller than the bandwidth of the intrinsic noise. Note that this relation does not follow from the expressions in the previous subsection, since they were derived by neglecting terms not proportional to the square of the local oscillator amplitude $\bar{a}_\mathrm{lo}$. For photodetectors with sufficiently small electrical noise, one can verify that the background noise floor vanishes when the lasers are turned off and that the prediction in \eqref{eq:Snolaser} is wrong. In other words, a photodetector is not affected by the intrinsic noise of the electromagnetic field in the absence of external light sources. We must then conclude that \eqref{eq:classnoiseDef3} is invalid.

Provided that such a noise characterization of the photodetector can be used to rule out intrinsic classical noise, it follows that classical noise in the electromagnetic field must originate from the lasers themselves or from the motion of the mechanical oscillator. Nevertheless, we know from experiments that even when removing the mechanical oscillator and filtering all noise from the lasers, the noise floor $S^{(o)}[\omega]$ will not vanish. This means that we need a model in which the noise floor only appears when the lasers are turned on. We will in the following consider different ways to achieve this.

Finally, we note that while a classical model for the intrinsic electromagnetic noise can be experimentally ruled out in the optical regime for photodetectors of sufficiently high quality, this might not be feasible in the microwave regime. Thus, one may have to entertain the possibility of classical noise in some situations. We will return to this issue in Section \ref{sec:ClassAsymm}.

\section{Quantum field and semiclassical detector}
\label{sec:QuantClass}

Our first example of a model that correctly describes the photocurrent noise floor is one where the electromagnetic field is quantized. This means that the intrinsic noise variables $\xi_\ext,\xi_\innt, \xi_\mathrm{lo}$ are now quantum operators. Our detector model will reflect that photocurrent is only generated when a photon arrives at the detector. However, we will still consider all the noise in the photocurrent to originate from the noise in the electromagnetic field. We therefore refer to this detector model as semiclassical, as it does not take into account the quantum nature of the interaction process in the photomultiplier. 

\subsection{Electromagnetic field noise}

In a quantum theory for the electromagnetic field, we can use the same expressions as before for the modes $a_\mathrm{out}$, $a_\mathrm{lo}$, and $a_\mathrm{det}$. The only change we need is to replace Equation \eqref{eq:classnoiseDef3} with \cite{Gardiner1985PRA,Clerk2010RMP}
\begin{align}
\label{eq:quantnoiseDef}
\langle \xi_i[\omega] \xi_j^\dagger[\omega'] \rangle & =  2 \pi \delta_{i,j} \delta(\omega + \omega') \notag \\
\langle \xi_i^\dagger[\omega] \xi_j[\omega'] \rangle & =  0 \\
\langle \xi_i[\omega] \xi_j[\omega'] \rangle & =  0 , \notag
\end{align}
to specify the properties of the quantum vacuum noise variables $\xi_\ext$, $\xi_\innt$, and $\xi_\mathrm{lo}$. We also note that the commutation relations
\begin{equation}
\label{eq:CommRelxi}
\big[\xi_i[\omega] , \xi^\dagger_j[\omega']\big] = 2\pi \delta_{ij} \delta(\omega + \omega')
\end{equation}
must hold.

\subsection{Detector model}

We now deviate from the classical square-law detector model in \eqref{eq:iClass}. This is based on the fact that a photoelectron in a photomultiplier is only emitted when energy is absorbed from the electric field \cite{Glauber1963PR}. We will go into more details on this in Section \ref{sec:ClassQuant}, but for now we simply let the average photocurrent be $\overline{i(t)} = \langle I(t) \rangle $, but with the definition
\begin{equation}
\label{eq:iadetordered}
I(t) =  q v^2  a^\dagger_\mathrm{det}(t) a_\mathrm{det}(t) .
\end{equation}
Crucially, the operators have not been symmetrized as in \eqref{eq:iadet}, such that in the absence of light sources, i.e., in the absence of photons, we get no photocurrent. Note, however, that in the limit of a strong local oscillator, Equation \eqref{eq:iadet2} is still valid. 

In order to calculate the photocurrent spectrum $S[\omega]$, we will still assume that the interpretation of $\overline{i(t+\tau) i(t)}$ in Equation \eqref{eq:iiCorrClass} is valid when defining $I(t)$ as in \eqref{eq:iadetordered}. It is worth noting that this does not {\it a priori} follow from the standard photodetection theory for a photomultiplier where all correlation functions are necessarily normal ordered and time ordered \cite{Glauber1963PR,Carmichael1987JOptSocAmB}. We return to this issue in Section \ref{sec:ClassQuant}. 
 
\subsection{Heterodyne spectrum and sideband asymmetry}

As before, we find that the noise spectrum of the photocurrent can be divided into three terms as in Eq.~\eqref{eq:Sclassclass}. The noise floor $S^{(o)}[\omega]$ now becomes the same as in \eqref{eq:Sclassclass-o}, but with $\alpha$ replaced by 1:
\begin{align}
\label{eq:Squantclass-o}
S^{(o)}[\omega] & = |Z|^2 \left[1  + \frac{r^2 \kappa}{4 \kappa_\ext} |\kappa_\ext \chi_c[-\tilde{\omega}] - \lambda|^2 \left(C_{xx} + C_{yy} \right) \right] .  
\end{align}
In the absence of laser noise, the noise floor in this model comes about from beating between the quantum vacuum noise of the electromagnetic field and the coherent local oscillator beam. However, unlike in Section \ref{sec:ClassClass}, the noise floor now vanishes in the absence of laser beams, due to the asymmetric properties \eqref{eq:quantnoiseDef} of the quantum vacuum noise operators.

The optomechanical term now becomes
\begin{align}
\label{eq:Squantclass-om}
S^{(om)}[\omega] & = - |Z|^2 \sqrt{\kappa_\ext} \\
& \times \mathrm{Im} \big(G^\ast \chi_c^\ast[-\tilde{\omega}] \big(S_{x\Gamma}[\tilde{\omega}] + S_{\Gamma x}[-\tilde{\omega}] \big) \big) , \notag
\end{align}
whereas the last term $S^{(m)}[\omega]$ is the same as in Eq.~\eqref{eq:Sclassclass-m}. This means that, as before, any sideband asymmetry at $\Delta = 0$ must be ascribed to optomechanical correlations, but now either due to quantum vacuum noise of the electromagnetic field or to classical laser noise.

Laser noise can in principle be removed prior to the beam reaching the optomechanical cavity. This was for example done in the experiment reported in Ref.~\cite{Underwood2015PRA}. If one can verify in some independent way that this has been achieved, the operator $\Gamma[\omega] = \Gamma_q[\omega]$ with $\Gamma_q[\omega]$ only given by quantum vacuum noise:
\begin{equation}
\label{eq:GammaQuant}
\Gamma_q[\omega] = \sqrt{\kappa_\ext \kappa_\innt} \chi_c[\omega] \xi_\innt[\omega] + \left( \kappa_\ext \chi_c[\omega] - 1 \right) \xi_\ext[\omega] .
\end{equation}
This means that the sideband asymmetry function $\Delta S[\tilde{\omega}]$, defined in Equation \eqref{eq:DeltaSDef}, then becomes
\begin{align}
\label{eq:DeltaSquantclass}
\Delta S[\tilde{\omega}] & = -|Z|^2 \sqrt{\kappa_\ext} G \\
& \times  \mathrm{Im} \big(\chi_c[\omega_m] S_{\Gamma_q x}[-\tilde{\omega}] - \chi_c^\ast[\omega_m] S_{\Gamma_q x}[\tilde{\omega}] \big) \notag
\end{align}
at $\Delta = 0$, since $S_{x \Gamma_q}[\tilde{\omega}] = 0$ due to the properties of the vacuum noise operators. In the end, we find that for the sideband asymmetry to be nonzero, the correlation functions
\begin{equation}
\label{eq:quantomcorr}
\langle \xi_\ext(\tau) x(0) \rangle \ , \ \langle \xi_\innt(\tau) x(0) \rangle 
\end{equation}
would have to be nonzero for at least some range of times $\tau$ (negative $\tau$ due to causality). This means that the oscillator position $x$ cannot possibly be considered a classical variable. If it were classical, one could always commute $x$ and $\xi_\ext$ or $\xi_\innt$, but then the expectation values in \eqref{eq:quantomcorr} would have to be zero, again due to the properties \eqref{eq:quantnoiseDef} of the vacuum noise operators.

\section{Classical field and quantum detector}
\label{sec:ClassQuant}

In this Section, we will consider the electromagnetic input field to be classical with no intrinsic noise. This should not at all be viewed as a claim that the electromagnetic field is actually classical, but it is done in the spirit of using the simplest possible model that can be consistent with the measurement results of the particular experiment we are describing. We will now use a different detector model, which takes into account the randomness associated with the emission of photoelectrons in the photomultiplier. This means that photocurrent noise does not necessarily originate from the noise in the electromagnetic field.

\subsection{Electromagnetic field noise}

Let us now assume that there is no intrinsic noise in the electromagnetic field, neither classical nor quantum. We still include classical laser noise. In other words, we use the noise properties from Section \ref{sec:ClassClass}, but with $\alpha = 0$.   

\subsection{Detector model}

We will now describe the detector according to standard photodetection theory for a photomultiplier \cite{Glauber1963PR,Carmichael1987JOptSocAmB}. Even though we will treat the electromagnetic input field classically in this section, we again allow for the possibility that the oscillator can be nonclassical. In order to emit a photoelectron via the photoelectric effect, energy must be absorbed from the electromagnetic field at the detector. This means that a simple perturbative calculation must give a transition amplitude $\mu^{(1)}_{i,f}(\mathbf{r},t)$ for emitting a photoelectron at position $\mathbf{r}$ and time $t$ proportional to the positive frequency part of the electromagnetic field, giving
\begin{equation}
\label{eq:AmplEmit}
\mu^{(1)}_{i\rightarrow f}(\mathbf{r},t) \propto \langle f | E^{(+)}(\mathbf{r},t) | i \rangle 
\end{equation}
where $E^{(+)}(\mathbf{r},t)$ is defined as in Equation \eqref{eq:EoutputPlus2}. We let $|i\rangle$ and $|f\rangle$ be the states of the mechanical oscillator's intrinsic environment prior to and after the photoelectron emission, respectively. If this mechanical bath is also classical, the {\it bra} and {\it ket} in Equation \eqref{eq:AmplEmit} can simply be ignored. The average rate $R^{(1)}(t)$ at which photoelectrons are generated is found by squaring this and summing over all final states:
\begin{align}
\label{eq:RateEmit}
R^{(1)}(t) & = \int d \mathbf{r} \, f(\mathbf{r}) \sum_f  |\mu^{(1)}_{i \rightarrow f}(t)|^2\\
 & \propto \int d \mathbf{r} \, f(\mathbf{r}) \langle i | E^{(-)}(\mathbf{r},t) E^{(+)}(\mathbf{r},t)| i \rangle \notag \\
 & \approx v^2 \langle i | a^\dagger_\mathrm{det}(t) a_\mathrm{det}(t) | i \rangle . \notag
\end{align}
The derivation of the last line is shown in Appendix \ref{app:Edet}. Since the electromagnetic field can have laser noise, and the mechanical bath is not necessarily in a pure state $|i\rangle$, this expectation value should be generalized to 
\begin{equation}
\label{eq:R1Result}
R^{(1)}(t) \propto \langle a^\dagger_\mathrm{det}(t) a_\mathrm{det}(t)  \rangle ,
\end{equation}
where $\langle \cdots \rangle$ is an ensemble average over both the classical electromagnetic noise and the noise in the mechanical bath. This justifies the expression for the operator $I(t)$ in Equation \eqref{eq:iadetordered}. 

Similarly, one finds that the transition amplitude for emitting two photoelectrons, one at time $t$ and position $\mathbf{r}$ and another at a later time $t + \tau > t$ at position $\mathbf{r'}$, is proportional to
\begin{equation}
\label{eq:AmplEmit2}
\mu^{(2)}_{i\rightarrow f}(\mathbf{r},t;\mathbf{r'},t+\tau) \propto \langle f | E^{(+)}(\mathbf{r'},t + \tau) E^{(+)}(\mathbf{r},t) | i \rangle .
\end{equation}
The probability per unit time squared of emitting two photoelectrons at times $t$ and $t + \tau > t$ must therefore be given by
\begin{align}
\label{eq:Prob2}
& R^{(2)}(t,t+\tau) \\
& = \int d \mathbf{r}  \int d \mathbf{r'} f(\mathbf{r}) f(\mathbf{r'}) \sum_f |\mu^{(2)}_{i\rightarrow f}(\mathbf{r},t;\mathbf{r'},t+\tau)|^2 \notag \\ 
& \propto \int d \mathbf{r}  \int d \mathbf{r'} f(\mathbf{r}) f(\mathbf{r'}) \notag \\
& \times \langle i | E^{(-)}(\mathbf{r},t) E^{(-)}(\mathbf{r'},t + \tau) E^{(+)}(\mathbf{r'},t + \tau) E^{(+)}(\mathbf{r},t) |i \rangle  \notag\\
& \approx v^4 \langle i | a^\dagger_\mathrm{det}(t) a^\dagger_\mathrm{det}(t+\tau) a_\mathrm{det}(t + \tau) a_\mathrm{det}(t) | i \rangle  \notag  \\
& \rightarrow v^4  \langle  a^\dagger_\mathrm{det}(t) a^\dagger_\mathrm{det}(t+\tau) a_\mathrm{det}(t + \tau) a_\mathrm{det}(t) \rangle . \notag
\end{align}
This suggests that any calculation of the noise properties of the photocurrent must involve expectation values where the electromagnetic field operators are both normal ordered and time ordered. In a completely classical theory, this ordering will of course have no effect. It is however important to note that we cannot commute $a^\dagger_\mathrm{det}$ and $a_\mathrm{det}$ when we allow for the possibility of a nonclassical mechanical oscillator.

We have so far ignored the fact that the emission of a photoelectron typically leads to a current pulse with a nonzero duration $\tau_d$. In other words, our assumption of infinite detector bandwidth means that we have assumed $\tau_d \rightarrow 0$. Let us now for a moment imagine that $\tau_d$ is finite and consider how to express the photocurrent autocorrelation 
\begin{equation}
\label{eq:iProd}
\overline{i(t+\tau) i(t)} 
\end{equation}
in terms of ensemble averages of optical and mechanical degrees of freedom. One possibility for the product $i(t+\tau) i(t)$ to be nonzero is if the two times $t$ and $t + \tau$ fall within the duration of two different photoelectric pulses. However, we can also get a contribution to this product from a single current pulse, as long as the two times fall within the duration of one pulse, i.e., if $|\tau| < \tau_d$. 

Taking into account these two possibilities, one arrives at the expression \cite{Carmichael1987JOptSocAmB} 
\begin{equation}
\label{eq:PhotocurrentCorrNT}
\overline{i(t+\tau) i(t)}  = \langle : I(t + \tau) I(t) : \rangle + q v^2 \delta(\tau) \langle I(t) \rangle 
\end{equation}
after taking the limit $\tau_d \rightarrow 0$. The colons indicate normal and time ordering and $I(t)$ is defined as in Equation \eqref{eq:iadetordered}. For $\tau > 0$, the first term in the parantheses becomes 
\begin{align}
\label{eq:PhotocurrentCorrNT2}
& \langle : I(t + \tau) I(t) : \rangle \\
& =  (qv^2)^2 \langle a_\mathrm{det}^\dagger(t) a_\mathrm{det}^\dagger(t+\tau) a_\mathrm{det}(t + \tau) a_\mathrm{det}(t) \rangle  \notag
\end{align}
when expressed in terms of the detector mode. The last term in \eqref{eq:PhotocurrentCorrNT} comes about due to self-correlation of photoelectric pulses, meaning that a single pulse can contribute to the product $i(t+\tau) i(t)$. 

Let us also point out that Equation \eqref{eq:PhotocurrentCorrNT} is correct also when using a quantum model for the electromagnetic input field. We have assumed that the detection efficiency is 1, which would mean that every photon that arrives at the detector generates a photoelectron. This approximation does not affect the interpretation of sideband asymmetry. It is however relevant when we compare different detector models in Section \ref{sec:Compare}, and we will comment further on this issue there.

\subsection{Heterodyne spectrum and sideband asymmetry}

We again express the photocurrent noise spectrum as in Equation \eqref{eq:Sclassclass}, i.e., in terms of a broadband noise floor $S^{(o)}[\omega]$, a term given by optomechanical correlations $S^{(om)}[\omega]$, and a term $S^{(m)}[\omega]$ given by the noise spectrum of the mechanical oscillator. In the limits \eqref{eq:AssumptionsBS}, the noise floor now becomes
\begin{equation}
\label{eq:Sclassquant-o}
S^{(o)}[\omega] = q v^2 i_0 + |Z|^2 \frac{r^2 \kappa}{4 \kappa_\ext} |\kappa_\ext \chi_c[-\tilde{\omega}] - \lambda|^2 \left(C_{xx} + C_{yy} \right)
\end{equation}
The noise floor in absence of laser noise is now proportional to the constant current $i_0$ as defined in Equation \eqref{eq:i0Def}. 

The first term of the noise floor \eqref{eq:Sclassquant-o} should now be interpreted as self-correlation of random photoelectric emissions whose rate of generation is proportional to the average photon flux impinging on the detector \cite{Carmichael1987JOptSocAmB}. In the strong local oscillator limit we work in, this flux is approximately given by the average flux of local oscillator photons and hence proportional to $\bar{a}_\mathrm{lo}^2$. The background noise in absence of laser noise is in other words not due to fluctuations already existing in the electromagnetic field and would be there even for a classical and noiseless field. The result $qv^2 i_0 = q^2 v^4 (1-T) \bar{a}_\mathrm{lo}^2$ in the first term of \eqref{eq:Sclassquant-o} should be compared to the equivalent term in \eqref{eq:Squantclass-o}, which was $|Z|^2 = q^2 v^4 T(1-T) \bar{a}_\mathrm{lo}^2$. The two results differ by a factor $T$, but are consistent within the approximation $1-T \ll 1$ that we have applied. The discrepancy can be traced back to the fact that we (justifiably) ignored the local oscillator quantum vacuum noise in the derivation of \eqref{eq:Squantclass-o}.
 
The present detector model also leads to a different optomechanical term in the photocurrent noise spectrum: 
\begin{equation}
\label{eq:Sclassquant-om}
S^{(om)}[\omega] = - 2 |Z|^2 \sqrt{\kappa_\ext} \, \mathrm{Im} \big(G^\ast \chi_c^\ast[-\tilde{\omega}] S_{x\Gamma}[\tilde{\omega}] \big) . 
\end{equation}
This is in fact the same as we had in the case of classical field {\it and} detector in Section \ref{sec:ClassClass}. This is natural, since only classical field noise can be directly detected with the present detector model. The optomechanical correlations that contribute to the spectrum can therefore only be a result of classical field noise.

Finally, the term $S^{(m)}[\omega]$ given by the oscillator noise spectrum now becomes
\begin{equation}
\label{eq:Sclassquant-m}
S^{(m)}[\omega] = |Z|^2 \kappa_\ext |G|^2 |\chi_c[-\tilde{\omega}]|^2 S_{xx}[\tilde{\omega}] .
\end{equation}
We should compare this to the result \eqref{eq:Sclassclass-m} that we obtained in Section \ref{sec:ClassClass} and Section \ref{sec:QuantClass}. The crucial difference is that $S^{(m)}[\omega]$ is now proportional to the mechanical oscillator spectrum $S_{xx}[\tilde{\omega}]$, {\it not} the symmetrized spectrum $\bar{S}_{xx}[\tilde{\omega}]$.

If we now define the sideband asymmetry function as before (Equation \eqref{eq:DeltaSDef}), consider detuning $\Delta = 0$, and assume no laser noise, we get
\begin{equation}
\label{eq:AsymmClassQuant}
\Delta S[\tilde{\omega}] = |Z|^2  \kappa_\ext |G|^2 |\chi_c[\omega_m]|^2 \left(S_{xx}[\tilde{\omega}] - S_{xx}[-\tilde{\omega}]\right)  . 
\end{equation}
We see that in this case, sideband asymmetry must originate from the quantum asymmetry between positive and negative frequencies in the mechanical oscillator noise spectrum.

\section{Quantum field and quantum detector}
\label{sec:QuantQuant}
It is well established that the electromagnetic field is indeed quantum, and that a photomultiplier emits photoelectrons only when photons are absorbed. With this knowledge, the most appropriate model to use in the optical regime seems to be one with quantum vacuum noise as defined in Section \ref{sec:QuantClass} and the detector model applied in Section \ref{sec:ClassQuant}. Conveniently, due to the normal and time ordering in Equation \eqref{eq:PhotocurrentCorrNT}, the expressions \eqref{eq:Sclassquant-o}, \eqref{eq:Sclassquant-om}, and \eqref{eq:Sclassquant-m} do not change when including the quantum vacuum noise. This is as expected for a photomultiplier - it does not "see" quantum vacuum noise but needs a real photon to react. 

However, even though the expressions are formally the same for a classical and a quantum input field when using the quantum detector model, that does not mean that the additional vacuum noise has no consequence. Unlike the photomultiplier, the mechanical oscillator is {\it not} blind to quantum vacuum noise and will be affected by it. This is famously referred to as radiation pressure shot noise \cite{Caves1980PRL} and will alter the mechanical oscillator spectrum $S_{xx}[\tilde{\omega}]$. We will see this explicitly in Section \ref{sec:MechModel}.

\section{Equivalence of detector models}
\label{sec:Compare}

We have introduced two different detector models for calculating the photocurrent spectrum $S[\omega]$. For convenience, we will refer to the semiclassical detector model used in Section \ref{sec:QuantClass} as the SCL model. The quantum detector model from Section \ref{sec:ClassQuant}, which led to normal and time ordered expectation values, will be referred to as the QUA model. In this Section, we will show that even though the photocurrent noise spectra calculated with the two different detector models appear to be quite different, they are in fact the same {\it if} we require the cavity output field to obey the same commutation relations as the input field \cite{Weinstein2014PRX}. 

\subsection{Consistent commutation relations}

In standard input-output theory \cite{Gardiner1985PRA,Clerk2010RMP}, the output variable $d_\mathrm{out}(t)$ necessarily obeys the same commutation relations as the input noise variables. However, this is not necessarily the case here, as we have taken Equation \eqref{eq:dFourier} as a starting point and allowed for the (somewhat strange) possibility that the cavity operator $d$ does not obey standard bosonic commutation relations. If we make the reasonable assumption that the output operator $d_\mathrm{out}(t)$ satisfies the same commutation relations as the input noise variables $\xi_\ext(t)$ and $\xi_\mathrm{int}(t)$, we have
\begin{equation}
\label{eq:adetComm}
\left[d_\mathrm{out}(t) , d_\mathrm{out}^\dagger(t') \right] = \delta(t - t') 
\end{equation}
and
\begin{equation}
\label{eq:adetComm2}
\big[d_\mathrm{out}(t) , d_\mathrm{out}(t') \big] = 0 . 
\end{equation}
The same commutation relations must then hold for the detector mode $a_\mathrm{det}(t)$. This can be used to rewrite the first term on the right hand side of Equation \eqref{eq:PhotocurrentCorrNT}. It is sufficient to consider $\tau > 0$, in which case we have
\begin{align}
\label{eq:Rewrite}
& \langle : I(t + \tau) I(t) : \rangle \\
& =  (qv^2)^2 \langle a_\mathrm{det}^\dagger(t) a_\mathrm{det}^\dagger(t+\tau) a_\mathrm{det}(t + \tau) a_\mathrm{det}(t) \rangle  \notag \\
& = (qv^2)^2 \Big[\langle a^\dagger_\mathrm{det}(t+\tau) a_\mathrm{det}(t+\tau) a_\mathrm{det}^\dagger(t) a_\mathrm{det}(t) \rangle \notag \\
& \ \quad \qquad - \delta(\tau)  \langle a^\dagger_\mathrm{det}(t) a_\mathrm{det}(t) \rangle \Big] \notag \\
& = \langle I(t+\tau) I(t) \rangle - qv^2 \delta(\tau) \langle I(t) \rangle . \notag
\end{align}
This means that Equations \eqref{eq:iiCorrClass} and \eqref{eq:PhotocurrentCorrNT} are in fact the same given the assumptions \eqref{eq:adetComm} and \eqref{eq:adetComm2}.

Let us now briefly mention the differences that would arise if the photon detection efficiency $\sigma$ differed from unity. The first term in Equation \eqref{eq:PhotocurrentCorrNT} would then be proportional to an additional factor $\sigma^2$, whereas the last term, which is the result of self-correlation of single photoelectric pulses, would only be proportional to $\sigma$. It would then still be possible to have equality between the two detector models (Equations \eqref{eq:iiCorrClass} and \eqref{eq:PhotocurrentCorrNT}). However, it would require introducing a fictitious field in Section \ref{sec:QuantClass} (where the field was quantum and the detector semiclassical) in order to account for vacuum fluctuations associated with the nonunit detection efficiency \cite{Carmichael1987JOptSocAmB}.

To see how the equivalence of the two detector models affects the interpretation of the photocurrent spectral density, let us write out the left hand side of Equation \eqref{eq:adetComm}, but now in the Fourier domain. This gives
\begin{align}
\label{eq:CommS}
 \left[d_\mathrm{out}[\omega] , d_\mathrm{out}^\dagger[\omega'] \right]   & = \big[\Gamma_q[\omega] , \Gamma_q^\dagger[\omega'] \big]  \\
& - i \sqrt{\kappa_\ext} \left(G \chi_c[\omega] \big[x[\omega],\Gamma^\dagger_q[\omega'] \big] \right. \notag \\
& \left. - G^\ast \chi_c^\ast[-\omega'] \big[\Gamma_q[\omega] , x[\omega'] \big] \right) \notag  \\
& + \kappa_\ext |G|^2 \chi_c[\omega] \chi_c^\ast[-\omega'] \big[ x[\omega] , x[\omega'] \big] \notag
\end{align}
where $\Gamma_q[\omega]$ is the quantum part of $\Gamma[\omega]$ as defined in Equation \eqref{eq:GammaQuant}. It is straightforward to show that
\begin{equation}
\label{eq:GammaqComm}
\left[d_\mathrm{out}[\omega] , d_\mathrm{out}^\dagger[\omega'] \right] = \left[\Gamma_q[\omega] , \Gamma_q^\dagger[\omega'] \right] = 2\pi \delta(\omega + \omega') ,
\end{equation}
which means that the remaining terms on the right hand side of \eqref{eq:CommS} must sum to zero. If we now take the expectation value of Equation \eqref{eq:CommS} and integrate over $\omega'$, we arrive at the relation
\begin{equation}
\label{eq:Relation}
\frac{2 \, \mathrm{Im}\left(G^\ast \chi_c^\ast[\omega] S_{\Gamma_q x}[\omega] \right)}{\sqrt{\kappa_\ext} |G|^2 |\chi_c[\omega]|^2 } = S_{xx}[\omega] - S_{xx}[-\omega] .
\end{equation}
Equation \eqref{eq:Relation} shows that there is a close relation between the amount of back-action on the oscillator from the radiation pressure shot noise and the magnitude of the oscillator's quantum zero-point fluctuations. This is a general feature of linear quantum measurements \cite{Khalili2012PRA,Weinstein2014PRX}, and the relation is valid even for a weak measurement where the probe beam has negligible effect on the spectrum $S_{xx}[\omega]$. We can use Equation \eqref{eq:Relation} to rewrite the result found with the SCL model in Section \ref{sec:QuantClass} (for $\Gamma = \Gamma_q$), giving
\begin{align}
\label{eq:SpectQuantClass}
& S^{(om)}[\omega] + S^{(m)}[\omega] \\
& =  - |Z|^2 \sqrt{\kappa_\ext} \, \mathrm{Im} \left(G^\ast \chi^\ast_c[-\tilde{\omega}] S_{\Gamma_q x}[-\tilde{\omega}] \right) \notag \\
& \quad + |Z|^2 \kappa_\ext |G|^2 |\chi_c[-\tilde{\omega}]|^2 \bar{S}_{xx}[\tilde{\omega}]  \notag \\
& = |Z|^2 \kappa_\ext |G|^2 |\chi_c[-\tilde{\omega}]|^2 S_{xx}[\tilde{\omega}] , \notag
\end{align}
which is exactly the $S^{(m)}[\omega]$ in Equation \eqref{eq:Sclassquant-m} that we found with the QUA model in Section \ref{sec:ClassQuant}.

We conclude that the two detector models result in the same photocurrent spectral density {\it provided} the assumption of equal commutation relations for the output and input electromagnetic modes \cite{Weinstein2014PRX}. Note that if we {\it assume} that standard quantum theory is correct, the equality of these commutation relations can of course be deduced.

\subsection{Optical versus microwave detectors}

We have seen that the standard photodetection theory for photomultipliers leads to the QUA detector model. This does not necessarily mean that the same detector model applies in the microwave regime, as the detection process there is quite different from the optical regime. We are not aware of any rigorous derivation of how the heterodyne spectrum should be expressed in terms of expectation values of field operators for microwave systems, and it is beyond the scope of this article to attempt to derive such a relation. In general, we can at least conclude that a detector described by the SCL model would have to not only absorb photons, but also emit them. The reason is that a detector that only absorbs photons can never detect quantum vacuum noise.

\section{The mechanical oscillator and its environment}
\label{sec:MechModel}

\subsection{Model}

To derive explicit expressions for the sidebands and their asymmetry, we need a model for the mechanical mode. We let the mechanical part of the Hamiltonian be given by 
\begin{equation}
\label{eq:Hmech}
H_\mathrm{mech} = \frac{P^2}{2m} + \frac{1}{2} m \omega_m^2 (X - X_0)^2 
\end{equation}
for a harmonic oscillator with effective mass $m$. The momentum $P$ and position $X$ should be interpreted as operators in a quantum theory. As before, we use the dimensionless position fluctuation $x$ as defined in Equation \eqref{eq:xDimless}. The variable $x$ may again be expressed in terms of coefficients $c$ and $c^\ast$, defined by
\begin{align}
\label{eq:cDefs}
c + c^\ast & = x = \frac{X - X_0}{X_\mathrm{zpf}} \\
i(c^\ast - c)  & = \frac{P}{P_\mathrm{zpf}} .
\end{align}
In a quantum theory, the coefficients $c, c^\ast$ are replaced by annihilation and creation operators $c, c^\dagger$ that obey $[c,c^\dagger] = 1$. 

The mechanical oscillator is also coupled to an environment. This could include coupling to a continuum of mechanical modes in the mechanical support, but it could also include the coupling to another optical mode used to cool the mechanical oscillator. We will assume that the interaction with these environments can be described by coupling to a single bath of continuum modes, similar to what we did for the cavity mode. The coefficient $c$ must then obey the quantum Langevin equation
\begin{equation}
\label{eq:cEOM}
\dot{c} = - \left(\frac{\gamma_m}{2} + i \omega_m\right) c - i\left(G^\ast d + G d^\ast \right) + \sqrt{\gamma_m} \eta .
\end{equation}
The oscillator has an intrinsic linewidth $\gamma_m$ as a result of coupling to its bath. We adopt a white and Gaussian noise model for the noise variable $\eta(t)$, and we let
\begin{align}
\label{eq:etaDef}
\langle \eta[\omega] \eta^\dagger[\omega'] \rangle & =  2 \pi  (n_\therm + \beta) \delta(\omega - \omega') \\
\langle \eta^\dagger[\omega] \eta[\omega'] \rangle & = 2 \pi n_\therm  \delta(\omega - \omega')  . \notag
\end{align}
The thermal phonon number $n_\therm$ is defined as in Equation \eqref{eq:nTherm}. In a quantum theory, it is the average phonon number of the mechanical mode when $G = 0$. The temperature $T$ in Equation \eqref{eq:nTherm} is in other words the effective temperature characterizing the mechanical mode's intrinsic bath. We again note that its value can for example depend on the power of another laser beam used for optomechanical cooling. 

In standard quantum theory, the constant $\beta = 1$. However, we will keep the symbol $\beta$ here in order to keep track of the contribution from quantum noise in the mechanical bath, in the same way as was done in Ref.~\cite{Weinstein2014PRX}. We note that in a theory where the mechanical oscillator and its environment is considered classical, $\beta$ would be zero and $n_\therm = k_B T/(\hbar \omega_m)$. 

From Equation \eqref{eq:cEOM}, we can find the dimensionless position $x$. To simplify the expressions a bit, let us now assume that the detuning $\Delta = 0$. This gives
\begin{align}
\label{eq:xSolution}
x[\omega] & = \sqrt{\gamma_m}\left(\chi_m[\omega] \eta[\omega] + \chi_m^\ast[-\omega] \eta^\dagger[\omega]\right) \\
& - iG\chi_c[\omega] \left(\chi_m[\omega] - \chi_m^\ast[-\omega]\right)\left(d_\mathrm{in}[\omega] +  d_\mathrm{in}^\dagger[\omega]\right) \notag
\end{align}
in the Fourier domain when defining the mechanical susceptibility as in Equation \eqref{eq:chimDef}. We have used that $G$ is real and $\chi_c^\ast[-\omega] = \chi_c[\omega]$ when $\Delta = 0$. The variable $d_\mathrm{in}$ was defined in Equation \eqref{eq:dinDef}. Depending on our model of the electromagnetic field, it can be either classical (a complex number) or a quantum operator.

\subsection{Generalized model for the electromagnetic field noise}

In addition to keeping track of where the quantum noise in the mechanical bath contributes, it is also instructive to track the noise in the electromagnetic field. To this end, let us now modify the quantum noise model in Equations \eqref{eq:quantnoiseDef}, such that we instead have
\begin{align}
\label{eq:quantnoiseDefMod}
\langle \xi_i[\omega] \xi_j^\dagger[\omega'] \rangle & =   2 \pi \alpha \, \delta_{i,j} \delta(\omega + \omega') \notag \\
\langle \xi_i^\dagger[\omega] \xi_j[\omega'] \rangle & =  0 \\
\langle \xi_i[\omega] \xi_j[\omega'] \rangle & =  0 , \notag
\end{align}
but we must keep in mind that standard quantum theory requires $\alpha = 1$.

The model \eqref{eq:quantnoiseDefMod} should not be confused with the classical model for intrinsic electromagnetic noise given in Equations  \eqref{eq:classnoiseDef3}. However, it will prove convenient that we have chosen the same symbol $\alpha$ to parametrize both the classical and quantum noise models.

\subsection{Mechanical noise spectrum}

We can now use Equation \eqref{eq:xSolution} to calculate the mechanical oscillator noise spectrum in the limits $|G| \ll \kappa$ and $\gamma_m \ll \omega_m$, giving
\begin{equation}
\label{eq:SpectSxx}
S_{xx}[\omega] = \gamma_m \left[ (\tilde{n}_\therm + \beta)|\chi_m[\omega]|^2 + \tilde{n}_\therm |\chi_m[-\omega]|^2 \right]  .
\end{equation}
Comparing with Equation \eqref{eq:SpectThermal} (by letting $\beta = 1$), we see that this is the (normalized) spectrum of an oscillator in a thermal state. However, the average phonon number is not $n_\therm$, but
\begin{equation}
\label{eq:ntildetherm}
\tilde{n}_\therm = n_\therm + p (\alpha + r C_{xx})
\end{equation}
where we define the dimensionless number
\begin{equation}
\label{eq:pDef}
p =  \frac{\kappa^2 |\chi_c[\omega_m]|^2}{4} C
\end{equation}
in terms of the cooperativity
\begin{equation}
\label{eq:Coop}
C \equiv \frac{4 G^2}{\kappa \gamma_m}  .
\end{equation}
The last term in Equation \eqref{eq:ntildetherm} is due to the fact that the probe beam noise leads to additional fluctuations in the mechanical oscillator position. For this additional heating to be negligible (in the sense that $\tilde{n}_\therm - n_\therm \ll 1$), we need $p (\alpha + r C_{xx}) \ll 1$. This is always satisfied if the amplitude laser noise is below shot noise level, i.e., $r C_{xx} \lesssim \alpha = 1$, and if the probe beam is sufficiently weak such that $p \ll 1$. On the other hand, we should note that the parameter $p$ cannot be too small. The reason is, as we will see below, that the signal to noise ratio of the sidebands also depends on this parameter.  

We observe that $S_{xx}[\omega]$ consists of two Lorentzians centered at $\pm \omega_m$. The asymmetry between these two peaks is for zero detuning only given by $\beta$, which means that it originates from the quantum nature of the mechanical oscillator's intrinsic bath. We emphasize that this conclusion and the expressions above are limited to the case $\Delta = 0$. For nonzero detuning, the probe beam can also give rise to modifications of the damping rate $\gamma_m$ and the mechanical frequency $\omega_m$, and the asymmetry is generally a result of both mechanical and optical quantum noise. 

Let us also write down the symmetrized noise spectrum
\begin{equation}
\label{eq:SxxbarExpl}
\bar{S}_{xx}[\omega] = \gamma_m (\tilde{n}_\therm + \beta/2)\left( |\chi_m[\omega]|^2 +  |\chi_m[-\omega]|^2   \right) ,
\end{equation}
since this enters our expression for the photocurrent spectrum when using the SCL detector model. It is also worthwhile to point out that the expression for the average phonon number \eqref{eq:ntildetherm} is unchanged if we replace the quantum noise model \eqref{eq:quantnoiseDefMod} with the classical model \eqref{eq:classnoiseDef3} for the intrinsic electromagnetic field noise.

\section{Explicit expressions with different detector models} 
\label{sec:Explicit}
In this section, we calculate the explicit expressions for the sidebands in the photocurrent spectrum using the two different detector models. We use the quantum vacuum noise model for the electromagnetic field in this section. Laser noise will be neglected from now on, assuming that this can be sufficiently suppressed by filtering. 

We will keep the parameters $\alpha$ and $\beta$ introduced in Equations \eqref{eq:quantnoiseDefMod} and \eqref{eq:etaDef} to keep track of where the quantum noise of the electromagnetic and mechanical baths contribute, similarly to what was done in Ref.~\cite{Weinstein2014PRX}. The standard quantum model is recovered by setting $\alpha = \beta = 1$. 

\subsection{Semiclassical detector model}
\label{sec:SCL}

Let us first use the SCL detector model presented in Section \ref{sec:QuantClass}. By using the expression \eqref{eq:Squantclass-om}, 
we find the optomechanical part of the photocurrent spectrum at frequency $\omega = \omega_\mathrm{if} + \tilde{\omega}$ to be
\begin{equation}
\label{eq:SomExplQuantClass}
S^{(om)}[\omega] = 2 \alpha \, p \, \bar{\kappa}_\ext  |Z|^2   \left( L[\tilde{\omega}] - L[-\tilde{\omega}] \right) ,
\end{equation}
introducing the quantity
\begin{equation}
\label{eq:kappabar}
\bar{\kappa}_\ext = \frac{\kappa_\ext}{\kappa} \leq 1.
\end{equation}
We have also defined the Lorentzian
\begin{equation}
\label{eq:Lpm}
L[\tilde{\omega}] = \frac{\left(\gamma_m/2\right)^2}{\left(\gamma_m/2\right)^2 + \left(\tilde{\omega} - \omega_m \right)^2} 
\end{equation}
such that its peak value is 1. We see that $S^{(om)}[\omega]$ is antisymmetric around $\omega_\imf$. Moving on to the mechanical term $S^{(m)}[\omega]$ which is given by Equation \eqref{eq:Sclassclass-m} with the SCL detector model, we find
\begin{align}
\label{eq:SmExplQuantClass}
S^{(m)}[\omega] & = 4 \left(\tilde{n}_\therm + \beta/2 \right)  \, p \,  \bar{\kappa}_\ext  |Z|^2 \left( L[\tilde{\omega}] + L[-\tilde{\omega}] \right) . 
\end{align}
As we have pointed out before, the contribution $S^{(m)}[\omega]$ is symmetric around $\omega_\imf$ with this detector model. 

\begin{figure}[htbp]
\begin{center} 
\includegraphics[width=.49\textwidth]{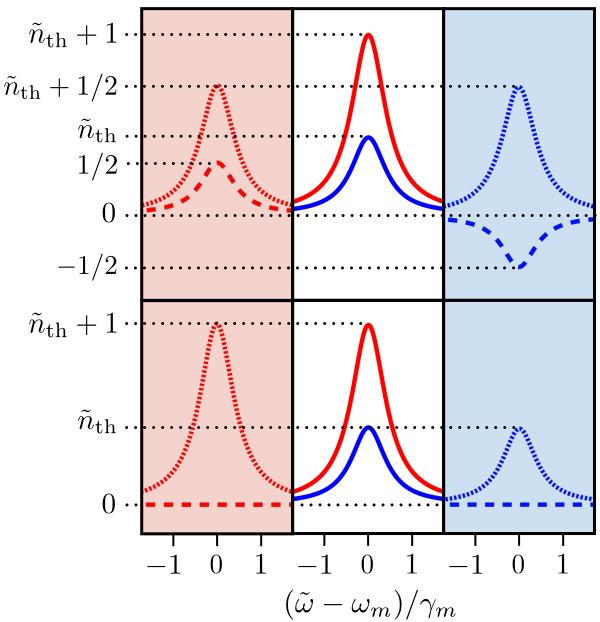}
\caption{Origin of sideband asymmetry with the two different detector models in absence of classical noise in the electromagnetic field. {\it Upper panels:} The SCL detector model. {\it Lower panels:} The QUA detector model. {\it Left:} The optomechanical contribution $S^{(om)}[\omega_\imf +\tilde{\omega}]$ (dashed) and the mechanical contribution $S^{(m)}[\omega_\imf + \tilde{\omega}]$ (dotted) at the red sideband. {\it Right:} The optomechanical contribution $S^{(om)}[\omega_\imf -\tilde{\omega}]$ (dashed) and the mechanical contribution $S^{(m)}[\omega_\imf - \tilde{\omega}]$ (dotted) at the blue sideband. {\it Center:} The sidebands with the noise floor subtracted. The red sideband $S_{rr}[\tilde{\omega}] - |Z|^2$ is taller than the blue, $S_{bb}[\tilde{\omega}] - |Z|^2$. All functions are presented in units of $4p\bar{\kappa}_\ext |Z|^2$.}
\label{fig:SCL}
\end{center}
\end{figure}

From these expressions, we can find the photocurrent spectrum at the red and blue sidebands as defined in Equations \eqref{eq:SrrDef} and \eqref{eq:SbbDef}:
\begin{align}
\label{eq:SrrSbbQuantClass}
S_{rr}[\tilde{\omega}] & = |Z|^2 \Big[\alpha + 4 \, p  \, \bar{\kappa}_\ext (\tilde{n}_\therm + \beta/2 + \alpha/2) L[\tilde{\omega}] \Big] \\
S_{bb}[\tilde{\omega}] & = |Z|^2 \Big[\alpha + 4 \, p  \, \bar{\kappa}_\ext (\tilde{n}_\therm + \beta/2 - \alpha/2) L[\tilde{\omega}] \Big] .
\end{align}
The upper panel of Figure \ref{fig:SCL} shows the two sidebands for $\alpha = \beta = 1$ as well as the individual terms in Equations \eqref{eq:SomExplQuantClass} and \eqref{eq:SmExplQuantClass}.
We can now calculate the sideband asymmetry function according to the SCL detector model. Using Equation \eqref{eq:DeltaSDef}, we get
\begin{equation}
\label{eq:DeltaSSemiClass}
\Delta S[\tilde{\omega}] = 4\alpha \, p  \, \bar{\kappa}_\ext |Z|^2 L[\tilde{\omega}] 
\end{equation}
 for $\tilde{\omega} \sim \omega_m$. Since this is proportional to $\alpha$ (and not $\beta$), it is explicitly clear that with this detector model, the sideband asymmetry comes entirely from the oscillator's response to quantum noise in the electromagnetic field, not from quantum noise in the oscillator's intrinsic bath.

\subsection{Quantum detector model}

If we instead use the QUA detector model from Section \ref{sec:ClassQuant} involving normal and time ordered expectation values, we find\begin{equation}
\label{eq:SomExplClassQuant}
S^{(om)}[\omega] = 0 .
\end{equation}
In other words, there is no contribution from optomechanical correlations with this detector model, as long as the electromagnetic field only has quantum vacuum noise. Using \eqref{eq:Sclassquant-m} to calculate the contribution $S^{(m)}[\omega]$, we now find
\begin{align}
\label{eq:SmExplClassQuant}
S^{(m)}[\omega] & = 4 \, p \,  \bar{\kappa}_\ext  |Z|^2  \left[(\tilde{n}_\therm + \beta) L[\tilde{\omega}] + \tilde{n}_\therm L[-\tilde{\omega}] \right] . 
\end{align}
We see that, unlike the expression in Equation \eqref{eq:SmExplQuantClass}, this is not symmetric around $\omega_\imf$. 

It is then straightforward to find the photocurrent spectrum at the red and blue sidebands:
\begin{align}
\label{eq:SrrSbbClassQuant}
S_{rr}[\tilde{\omega}] & = |Z|^2 \Big[1 + 4 \, p  \, \bar{\kappa}_\ext (\tilde{n}_\therm + \beta) L[\tilde{\omega}] \Big] \\
S_{bb}[\tilde{\omega}] & = |Z|^2 \Big[1 + 4 \, p  \, \bar{\kappa}_\ext \tilde{n}_\therm  L[\tilde{\omega}] \Big] .
\end{align}
The lower panel of Figure \ref{fig:SCL} shows how these two sidebands come about with the QUA detector model. The sideband asymmetry function becomes
\begin{equation}
\label{eq:DeltaSQuantQUA}
\Delta S[\tilde{\omega}] = 4 \beta \, p  \, \bar{\kappa}_\ext |Z|^2 L[\tilde{\omega}]  ,
\end{equation}
which is proportional to $\beta$, not $\alpha$. We thus conclude that with the QUA model and the assumption of zero detuning, the sideband asymmetry originates entirely from quantum vacuum noise in the oscillator's intrinsic bath.

\subsection{Discussion}

To recover standard quantum theory, we should set $\alpha = \beta = 1$.
As expected, we then see that the two detector models give the same result. They both give the red and blue sideband spectra
\begin{align}
\label{eq:SrrSbbStandard}
S_{rr}[\tilde{\omega}] & = |Z|^2 \Big[1 + 4 \, p  \, \bar{\kappa}_\ext (\tilde{n}_\therm + 1) L[\tilde{\omega}] \Big] \\
S_{bb}[\tilde{\omega}] & = |Z|^2 \Big[1 + 4 \, p  \, \bar{\kappa}_\ext \tilde{n}_\therm  L[\tilde{\omega}] \Big] \notag
\end{align}
as well as the asymmetry function
\begin{equation}
\label{eq:DeltaSQuant}
\Delta S[\tilde{\omega}] = 4 \, p  \, \bar{\kappa}_\ext |Z|^2 L[\tilde{\omega}]  .
\end{equation}
We note that the peak value of $\Delta S[\tilde{\omega}]$ relative to the flat background noise $S^{(o)}[\omega] \equiv S^{(o)}$ is
\begin{equation}
\label{eq:DeltaSRatio}
\frac{\Delta S[\omega_m]}{S^{(o)}} = 4 \, p  \, \bar{\kappa}_\ext .
\end{equation}
In other words, the difference between the heights of the red and blue sidebands in units of the noise floor is given by the right hand side of \eqref{eq:DeltaSRatio}. We note that the right hand side is not explicitly dependent on the temperature of the mechanical mode and can be determined by independent measurements.

We nevertheless see that the asymmetry comes about for different reasons with the two different detector models, as shown graphically in Figure \ref{fig:SCL}. A probe at zero detuning will measure properties of the oscillator's intrinsic bath with a detector described by the QUA model. A detector described by the SCL model will on the other hand measure how the oscillator responds to the noise in the electromagnetic field. While these are indistinguishable in standard quantum theory, one could perhaps imagine that they would differ in some modified theory. We will not speculate on such theories here, but it is worth pointing out that searching for or ruling out deviations from standard quantum theory is in fact one of the major goals in the field of optomechanics. At the very least, the above shows that detailed knowledge of the measurement process can be very important when doing so. 

%

Finally, we should point out that in the case where the oscillator is cooled close to the ground state by a separate laser beam, the properties of the noise variable $\eta$ may largely originate from quantum noise associated with another cavity mode and not from mechanical degrees of freedom. This means that even though the QUA detector model leads to a measurement of the properties of the intrinsic noise $\eta$, that does not necessarily mean that one measures properties of a mechanical bath. To be able to do that would require that the oscillator is sufficiently cooled by other means (i.e., not laser cooling) in order to resolve the asymmetry.

\section{Failure of a fully classical model}
\label{sec:ClassAsymm}

We have seen in Sections \ref{sec:QuantClass} and \ref{sec:SCL} that with the SCL detector model and a quantum electromagnetic field, the sideband asymmetry originates from optomechanical correlations due to quantum vacuum noise of the electromagnetic field. This raises two important questions: 1) Can the sideband asymmetry also be explained by intrinsic {\it classical} noise of the electromagnetic field? 2) If yes, does that mean that sideband asymmetry measurements on optomechanical systems can always be explained by classical theories with classical mechanical oscillators? In this Section, we will show that the answer to question 1) is indeed {\it yes}. The right type of classical field noise can give just the right amount of sideband asymmetry expected in a quantum theory. However, we will show that the answer to question 2) is {\it no}. A theory with a classical mechanical oscillator with no zero point motion {\it cannot} recreate the results of a quantum theory, provided that the oscillator is cooled down to sufficiently low temperatures. That being said, in Section \ref{sec:limitsLaserCooling} we show that this low temperature regime cannot be reached by laser cooling in the classical model, since the classical electromagnetic field noise sets a lower limit on the temperature that can be reached by this technique.

\subsection{Heterodyne spectrum in a classical model}

Let us now assume that the noise in the electromagnetic field is fully classical, such that $\xi_\ext$ and $\xi_\innt$ are classical, commuting variables. We again neglect laser noise, but we use the model for intrinsic classical noise given in Equation \eqref{eq:classnoiseDef3}. This of course assumes that we are not able to rule out such a model through a noise characterization of our detector. We also assume that the mechanical oscillator is classical, which means that we must set the parameter $\beta = 0$. 

With these assumptions, we can use the expressions from Section \ref{sec:ClassClass} to calculate the photocurrent spectrum which will depend on the parameter $\alpha$ as defined in Equation \eqref{eq:classnoiseDef3}. It turns out that the result is identical to what we found with a quantum electromagnetic field in Section \ref{sec:SCL}, as long as we replace $\beta$ with 0. The red and blue sideband spectra then become
\begin{align}
\label{eq:SrrSbbClassClass}
S_{rr}[\tilde{\omega}] & = |Z|^2 \Big[\alpha + 4 \, p  \, \bar{\kappa}_\ext \left(\tilde{n}_\therm + \alpha/2\right) L[\tilde{\omega}] \Big] \\
S_{bb}[\tilde{\omega}] & = |Z|^2 \Big[\alpha + 4 \, p  \, \bar{\kappa}_\ext \left(\tilde{n}_\therm - \alpha/2\right) L[\tilde{\omega}] \Big] , \notag
\end{align}
with $\tilde{n}_\therm = n_\therm + p \alpha$. Since we are now dealing with classical physics, we cannot think of $n_\therm$ as an average number of phonons, but we should rather make the identification
\begin{equation}
\label{eq:nthermClass}
n_\therm = \frac{k_B T}{\hbar \omega_m} ,
\end{equation}
where $T$ is the temperature characterizing the mechanical oscillator's intrinsic bath.

The noise floor is now $|\tilde{Z}|^2 \equiv |Z|^2 \alpha$. In an experiment, one would probably not know the value of neither $|Z|$ nor $\alpha$, so it is more meaningful to write the spectra in units of the noise floor. Taking $\alpha$ outside the parantheses allows us to express the spectra as
\begin{align}
\label{eq:SrrSbbClassClass2}
S_{rr}[\tilde{\omega}] & = |\tilde{Z}|^2 \Big[1 + 4 \, p  \, \bar{\kappa}_\ext \left(\tilde{n}_\mathrm{inf} + 1 \right) L[\tilde{\omega}] \Big] \\
S_{bb}[\tilde{\omega}] & = |\tilde{Z}|^2 \Big[1 + 4 \, p  \, \bar{\kappa}_\ext \tilde{n}_\mathrm{inf} L[\tilde{\omega}] \Big] \notag .
\end{align}
We see that by defining the parameter
\begin{equation}
\label{eq:ninf}
\tilde{n}_\mathrm{inf} = \frac{\tilde{n}_\therm}{\alpha} - \frac{1}{2} = \frac{n_\therm}{\alpha} +  p -  \frac{1}{2} ,
\end{equation}
the spectra \eqref{eq:SrrSbbClassClass2} take exactly the same form as in \eqref{eq:SrrSbbStandard}. We also see that the spectra satisfy Equation \eqref{eq:DeltaSRatio}, meaning that the height difference of the sidebands in units of the noise floor is just the same as in a quantum theory.

The only difference from the quantum result is that $\tilde{n}_\therm$ is replaced by $\tilde{n}_\mathrm{inf}$, but the mechanical mode's effective temperature is typically unknown anyway. This means that if the correct theory was in fact classical, but we mistakenly thought the quantum theory was correct, we would incorrectly infer that the average phonon number was $\tilde{n}_\mathrm{inf}$. 

\begin{figure}[htbp]
\begin{center} 
\includegraphics[width=.49\textwidth]{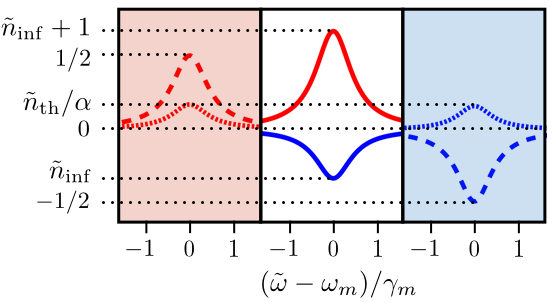}
\caption{Classical optomechanical correlations can lead to a negative blue sideband in the low temperature regime, i.e., noise squashing. This does not occur in the quantum theory, where the blue sideband height will asymptotically approach $p$ from above as the temperatur decreases. {\it Left:} The optomechanical contribution $S^{(om)}[\omega_\imf +\tilde{\omega}]$ (dashed) and the mechanical contribution $S^{(m)}[\omega_\imf + \tilde{\omega}]$ (dotted) at the red sideband. {\it Right:} The optomechanical contribution $S^{(om)}[\omega_\imf -\tilde{\omega}]$ (dashed) and the mechanical contribution $S^{(m)}[\omega_\imf - \tilde{\omega}]$ (dotted) at the blue sideband. {\it Center:} The sidebands with the noise floor subtracted. Here, the red sideband $S_{rr}[\tilde{\omega}] - |\tilde{Z}|^2$ is positive, whereas the blue $S_{bb}[\tilde{\omega}] - |\tilde{Z}|^2$ is negative. All functions are presented in units of $4p\bar{\kappa}_\ext |\tilde{Z}|^2$.}
\label{fig:NegBlue}
\end{center}
\end{figure}

There is, however, a crucial difference between $\tilde{n}_\therm$ and $\tilde{n}_\mathrm{inf}$. The temperature $T$ and thus $\tilde{n}_\therm$ must always be positive, but for sufficiently low temperatures and small $p$, one could reach $\tilde{n}_\therm/\alpha < 1/2$. This would result in a  negative $\tilde{n}_\mathrm{inf}$. In other words, for low temperatures
\begin{equation}
\label{eq:TBelow}
T < \frac{\hbar \omega_m \alpha}{k_B} \left(\frac{1}{2} - p \right) ,
\end{equation}
we could have $|S^{(om)}[\omega]| > S^{(m)}[\omega]$ at the blue sideband, which would lead to a negative Lorentzian (relative to the noise floor), rather than a positive one. This means that classical optomechanical correlations can lead to so-called noise squashing. See Figure \ref{fig:NegBlue} for an illustration. In contrast, in a quantum theory with the SCL detector model, the negative contribution from quantum optomechanical correlations is always compensated by the oscillator's zero point motion. 

\begin{figure}[htbp]
\begin{center} 
\includegraphics[width=.49\textwidth]{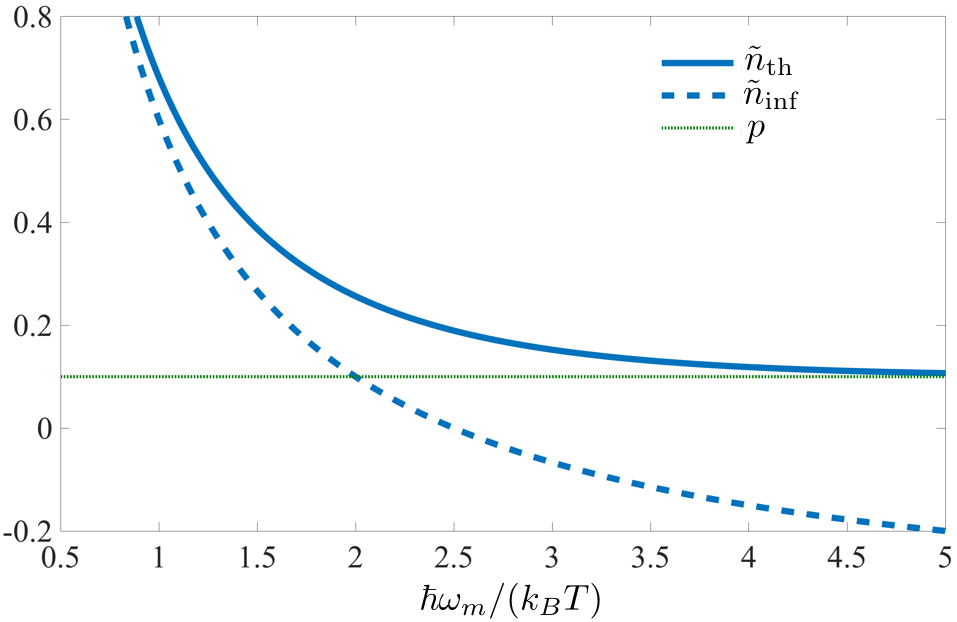}
\caption{The blue sideband height above the noise floor in units of $4 p \bar{\kappa}_\ext |Z|^2$ as a function of inverse temperature for the quantum model (solid curve) and the classical model (dashed curve). We have chosen $\alpha = 1$ and $p = 0.1$ here. We see that for all $T$ where $\tilde{n}_\mathrm{inf} > p$, the magnitude of the derivative of $\tilde{n}_\mathrm{inf}$ is above a finite value.}
\label{fig:BlueHeight}
\end{center}
\end{figure}

In the quantum theory, the height of the blue sideband is proportional to $\tilde{n}_\therm$ which can never go below the value $p$ because of radiation pressure shot noise. This means that as the temperature is lowered, the value $\tilde{n}_\therm$ will asymptotically approach $p$ from above. The classical theory does not have this feature. To be more specific, let us define the normalized inverse temperature
\begin{equation}
\label{eq:QDef}
Q = \frac{\hbar \omega_m}{k_B T} . 
\end{equation}
For all temperatures $T$ for which $\tilde{n}_\mathrm{inf} > p$, we then have 
\begin{equation}
\label{eq:dninfdq}
\left|\frac{\partial \tilde{n}_\mathrm{inf}}{\partial Q}\right| > \frac{\alpha}{4} .
\end{equation}
While $\alpha$ is generally unknown, one might be able to put a lower limit on its value. In practice, one would only be worried about the case $\alpha \geq 1$. This means that a lower limit can be placed on the absolute value of the derivative \eqref{eq:dninfdq} in a classical theory. In contrast, in the quantum theory, where $n_\therm$ is given by Equation \eqref{eq:nTherm}, there is no such limit and we have 
\begin{equation}
\label{eq:dnthTildedq}
\lim_{Q \rightarrow \infty} \left|\frac{\partial \tilde{n}_\mathrm{th}}{\partial Q}\right| = 0 .
\end{equation}
The behavior of the blue sideband height as a function of inverse temperature is depicted in Figure \ref{fig:BlueHeight} for both the quantum and the classical theory. This qualitative difference is in principle observable.  

We conclude that in the low temperature regime, the predictions from a completely classical theory do differ from those of a theory with a quantum mechanical oscillator. 

\subsection{Limits to laser cooling with classical field noise}
\label{sec:limitsLaserCooling}

Even though the classical and quantum theories differ at low temperatures, we will now see that laser cooling cannot be used to expose this deviation. Let us imagine that the mechanical oscillator is cooled with  the aid of another cavity mode. We let the cavity mode used for cooling have the same linewidth $\kappa$ and the same classical intrinsic noise properties \eqref{eq:classnoiseDef3} as the cavity mode used for measurement. The cooling cavity mode is driven by another laser at a detuning $\Delta_\cool$, and the interaction between this cavity mode and the mechanical oscillator is characterized by the many-photon optomechanical coupling $G_\cool$. We can disregard the measurement beam for now. 

We still consider a classical theory in which $\beta = 0$. Besides coupling to the cooling cavity mode, the oscillator also interacts with its mechanical support. Let us assume that the latter gives rise to an intrinsic mechanical linewidth $\gamma_{m,0}$. We also let the temperature of the mechanical support be $T_0$. The oscillator's resonance frequency before coupling to the cooling mode is $\omega_{m,0}$. Defining the cooling cavity mode susceptibility
\begin{equation}
\label{eq:chicool}
\chi_{c,2}[\omega] = \frac{1}{\kappa/2 - i(\omega + \Delta_\cool)} ,
\end{equation}
the standard linearized theory of optomechanics then gives the oscillator's effective linewidth
\begin{equation}
\label{eq:gammamEff}
\gamma_m = \gamma_{m,0} + 2 |G_\cool|^2 \mathrm{Re}   \left(\chi_{c,2}[\omega_m] - \chi_{c,2}^\ast[-\omega_m] \right)
\end{equation}
and effective resonance frequency
\begin{equation}
\label{eq:omegamEff}
\omega_m = \omega_{m,0} +  |G_\cool|^2 \mathrm{Im} \left(|\chi_{c,2}[\omega_m] - \chi_{c,2}^\ast[-\omega_m] \right)
\end{equation}
in the limit $\gamma_m \ll \kappa , \omega_m$. The effective temperature $T$ of the mechanical mode is given by the number $n_\therm = k_B T/\hbar \omega_m$, which becomes
\begin{equation}
\label{eq:nthEff}
n_\therm = \frac{\gamma_{m,0} n_{\therm,0}}{\gamma_m} + \frac{\alpha \kappa |G_\cool|^2  \left(|\chi_{c,2}[\omega_m]|^2 + |\chi_{c,2}[-\omega_m]|^2 \right)}{2\gamma_m} .
\end{equation}
Here, we have defined 
\begin{equation}
\label{eq:ntherm0}
n_{\therm,0} = \frac{k_B T_0}{\hbar \omega_{m}} ,
\end{equation}
which is the temperature of the mechanical bath in units of $\hbar \omega_m/k_B$.

For a sufficiently strong cooling laser, we can reach the regime $\gamma_m \gg \gamma_{0,m}$, in which case expression \eqref{eq:nthEff} can be written as
\begin{equation}
\label{eq:nthEff2}
n_\therm = \frac{\gamma_{m,0} n_{\therm,0}}{\gamma_m} + \frac{\alpha\left(|\chi_{c,2}[\omega_m]|^2 + |\chi_{c,2}[-\omega_m]|^2\right)}{2\left(|\chi_{c,2}[\omega_m]|^2 - |\chi_{c,2}[-\omega_m]|^2\right)}  .
\end{equation}
The second term in this expression arises from intrinsic field noise entering the cooling cavity. Unlike the first term, this cannot be made arbitrarily small. We find that the second term leads to
\begin{equation}
\label{eq:nthermLimitCooling}
n_\therm > \frac{\alpha}{2} , 
\end{equation}
or in terms of temperature,
\begin{equation}
\label{eq:TLimitCooling}
T > \frac{\hbar \omega_m \alpha}{2 k_B} . 
\end{equation}
Comparing this with Equation \eqref{eq:TBelow}, we see that when cooling with a cavity mode with the same classical noise strength as the measurement mode, one can never reach the desired regime where classical and quantum theories are distinguishable.

Let us now consider the resolved sideband regime $\omega_m \gg \kappa$ and the detuning $\Delta_\cool = - \omega_m$. Equation \eqref{eq:nthEff2} then becomes  
\begin{equation}
\label{eq:nthEff3}
n_\therm = \frac{\gamma_{m,0} n_{\therm,0}}{\gamma_m} + \alpha \left[ \frac{1}{2} + \left(\frac{\kappa}{4 \omega_m} \right)^2 \right] .
\end{equation}
This should be compared with the quantum result \cite{Marquardt2007PRL,Wilson-Rae2007PRL} 
\begin{equation}
\label{eq:nthEffQuant}
n_\therm = \frac{\gamma_{m,0} n_{\therm,0}}{\gamma_m}  + \left(\frac{\kappa}{4 \omega_m} \right)^2
\end{equation}
that follows from using the noise model \eqref{eq:quantnoiseDefMod} with $\alpha = 1$. Inserting Equation \eqref{eq:nthEff3} into the expression for $\tilde{n}_\mathrm{inf}$ in Equation \eqref{eq:ninf} gives
\begin{equation}
\label{eq:ninfLimit}
\tilde{n}_\mathrm{inf} = \frac{\gamma_{m,0} n_{\therm,0}}{\alpha \gamma_m} + \left(\frac{\kappa}{4 \omega_m} \right)^2 + p
\end{equation}
which is always positive. We emphasize that in a classical world with $\alpha = 1$, we would mistakenly measure an apparent average phonon number $\tilde{n}_\mathrm{inf}$ which is {\it exactly equal} to the average phonon number $\tilde{n}_\therm = n_\therm + p$ one finds in a quantum theory. If $\alpha \neq 1$, $\tilde{n}_\mathrm{inf}$ simply differs from the expected quantum result by $T_0 \rightarrow T_0/\alpha$.

In general, it is impossible to reach the temperature regime \eqref{eq:TBelow} by having the oscillator interact with degrees of freedom whose intrinsic classical noise is at least as large as that of the measurement cavity mode. However, for a large mechanical frequency, i.e., for an oscillator in the GHz regime, it can be possible to reach this temperature regime simply by conventional cooling of the oscillator's surroundings such that $T_0$ satisfies Equation \eqref{eq:TBelow}. In other words, the regime where classical and quantum theories for the heterodyne measurement spectrum differ can be accessible in cases where ground state cooling can be achieved through direct cryogenic cooling \cite{OConnell2010Nature,Meenehan2015PRX}.

\section{Conclusion}
\label{sec:Conclusion}
We have looked at the heterodyne photocurrent spectrum in the context of quantum optomechanical systems. The interpretation of sideband asymmetry in a quantum theory without classical noise is dependent on the choice of detector model and cannot be determined without detailed knowledge of the measurement process. For a detector that measures normal and time ordered expectation values, the sideband asymmetry reflects the quantum asymmetry in the oscillator's noise spectrum. For a detector that measures symmetrized expectation values, the asymmetry is a result of interference between quantum noise in the electromagnetic field and the position of the mechanical oscillator. Nevertheless, in both types of measurements, the sidebands are the same and the asymmetry reflects the nonclassical nature of the mechanical oscillator.

We have seen that sideband asymmetry can arise in a classical theory as well, irrespective of detector model, when the electromagnetic field has intrinsic classical noise. The asymmetry is then ascribed to interference between classical field noise and oscillator position. The magnitude of the asymmetry in units of the noise floor is in fact exactly the same as expected in a quantum theory. However, we have pointed out that at very low temperatures, the blue sideband can become negative relative to the noise floor in a classical theory. This noise squashing occurs when the magnitude of the interference term exceeds the contribution from the oscillator's noise spectrum. In a quantum theory with no classical noise, this is not possible, due to quantum zero point fluctuations of the mechanical oscillator. We have thus argued that while sideband asymmetry itself cannot determine whether the oscillator is classical or quantum, the height of the blue sideband as a function of temperature is qualitatively different in a quantum versus a classical theory. We have also emphasized that classical electromagnetic field noise can in some cases be ruled out by other means, such as filtering of laser noise and a proper characterization of the background noise seen by the photodetector.

\begin{acknowledgements}
The author would like to thank Konrad Lehnert, Aashish Clerk, Jack Harris, Alexey Shkarin, Steven Girvin, and Andreas Nunnenkamp for valuable input.
\end{acknowledgements}

\appendix

\section{Expressing the mechanical noise spectrum in terms of equal time expectation values}
\label{app:EqTime}

Let us start by using Equations \eqref{eq:EOMMechOsc} to derive differential equations for the correlation functions $\langle x(\tau) x(0) \rangle$ and $\langle p(\tau) x(0) \rangle$ for $\tau > 0$. This gives
\begin{align}
\label{eq:DiffEqCorr}
\partial_\tau \langle x(\tau) x(0) \rangle & = \omega_m \langle p(\tau) x(0) \rangle \\
\partial_\tau \langle p(\tau) x(0) \rangle & = - \omega_m \langle x(\tau) x(0) \rangle - \frac{\gamma_m}{2} \langle p(\tau) x(0) \rangle \notag
\end{align}
when using that $\langle F(\tau) x(0) \rangle = 0$ for $\tau > 0$. We now define the Laplace transforms
\begin{align}
\label{eq:Laplace}
{\cal X}(s) & = \int_0^\infty d \tau \, e^{-s\tau} \langle x(\tau) x(0) \rangle \\
{\cal P}(s) & = \int_0^\infty d \tau \, e^{-s\tau} \langle p(\tau) x(0) \rangle . \notag
\end{align}
Transforming the differential equations to the $s$-domain then gives 
\begin{align}
\label{eq:EOMLaplace}
s {\cal X}(s) - \langle x^2 \rangle & = \omega_m {\cal P}(s) \\
s {\cal P}(s) - \langle px \rangle & = - \omega_m {\cal X}(s) - \frac{\gamma_m}{2} {\cal P}(s) . \notag
\end{align}
We can solve these algebraically to find
\begin{equation}
\label{eq:XSol}
{\cal X}(s) = \frac{(\gamma_m/2 + s) \langle x^2 \rangle + \omega_m \langle p x \rangle}{(\gamma_m/2 + s)s + \omega_m^2} .
\end{equation}
In the limit $\gamma_m \ll \omega_m$, the transformation back to the time domain gives
\begin{equation}
\label{eq:xCorrSol}
\langle x(\tau) x(0) \rangle = e^{-\gamma_m \tau/2} \left[\cos(\omega_m \tau) \langle x^2 \rangle + \sin(\omega_m \tau) \langle p x \rangle   \right] . 
\end{equation}
We emphasize that this result is only valid for $\tau > 0$.

To find the same correlation function for $\tau < 0$, we can use $\langle x(\tau) x(0) \rangle = \langle x(0) x(\tilde{\tau}) \rangle$ with $\tilde{\tau} = -\tau > 0$, which follows from time translational symmetry. Through a similar calculation as above, we then find
\begin{equation}
\label{eq:xCorrSolNeg}
\langle x(\tau) x(0) \rangle = e^{\gamma_m \tau/2} \left[\cos(\omega_m \tau) \langle x^2 \rangle - \sin(\omega_m \tau) \langle x p \rangle  \right] . 
\end{equation}
We have again used causality, i.e., $\langle x(0) F(\tilde{\tau}) \rangle = 0$. 

We are then ready to calculate the noise spectrum, which can be written
\begin{equation}
\label{eq:SDivided}
S_{xx}[\omega] = \lim_{\varepsilon \rightarrow 0^+} \Big\{\int_\varepsilon^\infty + \int^{-\varepsilon}_{-\infty} + \int^{\varepsilon}_{-\varepsilon} \Big\} d \tau \, e^{i \omega \tau} \langle x(\tau) x(0) \rangle .
\end{equation}
For $\varepsilon \ll 2\pi/\omega_m$, the last term can be approximated by $2 \varepsilon \langle x^2 \rangle \rightarrow 0$. Using Equations \eqref{eq:xCorrSol} and \eqref{eq:xCorrSolNeg} to calculate the other terms leads to Equation \eqref{eq:SpectSingleTime} in the main text. Alternatively, one can use the Laplace transform ${\cal X}(s)$ to find $S_{xx}[\omega]$ directly by analytical continuation.

\section{Formalism for the electromagnetic field}
\label{app:Formalism}
In this section, we will briefly review the formalism used to describe the electromagnetic field. The motivation for this is to show the similarities and differences between a quantum and a classical description.

\subsection{Hamiltonian for a classical cavity field}

The Hamiltonian for the electromagnetic field inside the cavity is given by
\begin{equation}
\label{eq:ElmagH}
H_\mathrm{cav} = \frac{1}{2} \int dV \left(\varepsilon_0 \mathbf{E}^2(\mathbf{r},t) + \mu_0 \mathbf{B}^2(\mathbf{r},t) \right)
\end{equation}
where $\varepsilon_0$ is the vacuum electric permittivity, $\mu_0$ the vacuum magnetic susceptibility, and $\mathbf{E}(\mathbf{r},t)$ and $\mathbf{B}(\mathbf{r},t)$ the electric and magnetic fields.

The electric field inside the cavity can be written as an expansion in terms of real mode functions $\mathbf{u}_j(\mathbf{r})$,
\begin{equation}
\label{eq:Efield}
\mathbf{E}(\mathbf{r},t) = \frac{1}{\sqrt{\varepsilon_0}} \sum_j p_j(t) \mathbf{u}_j(\mathbf{r}) ,
\end{equation}
where $\varepsilon_0$ is the electric permittivity. If the space between the mirror contains no charges, we must have $\nabla \cdot \mathbf{u}_j(\mathbf{r}) = 0$. In order for the electric field to satisfy the wave equation, the orthonormal mode functions $\mathbf{u}_j(\mathbf{r})$ must satisfy
\begin{equation}
\label{eq:udef}
\nabla^2 \mathbf{u}_j(\mathbf{r}) = - k_j^2 \mathbf{u}_j(\mathbf{r}) 
\end{equation}
with constant $k_j$, and $p_j(t)$ are coefficients associated with these mode functions. The boundary conditions for the mode functions are $\hat{n}(\mathbf{r}) \times \mathbf{u}_j(\mathbf{r}) = 0$, where $\hat{n}$ is a vector normal to the mirror surface. 

The magnetic field can be expressed as
\begin{equation}
\label{eq:Bfield}
\mathbf{B}(\mathbf{r},t) = - \frac{1}{\sqrt{\varepsilon_0}} \sum_j q_j(t) \left(\nabla \times \mathbf{u}_j(\mathbf{r}) \right)
\end{equation}
where the coefficients $q_j$ must satisfy $\dot{q}_j = p_j$.

When inserting the fields in Eqs.~\eqref{eq:Efield} and \eqref{eq:Bfield} into the Hamiltonian \eqref{eq:ElmagH}, one arrives at
\begin{equation}
\label{eq:ElmagH2}
H_\mathrm{cav} = \frac{1}{2} \sum_j \left(p_j^2 + \omega_j^2 q_j^2 \right)
\end{equation}
with the angular frequencies defined as $\omega_j = c k_j$. We recognize this as the Hamiltonian for a collection of harmonic oscillators with mass $m = 1$. In other words, we can think of the coefficients $p_j$ as momenta per square root of mass, and the coefficents $q_j$ as positions times square root of mass.

We now define the complex coefficients
\begin{equation}
\label{eq:aDef}
a_j = \sqrt{\frac{\omega_j}{2K}} \left(q_j + \frac{i}{\omega_j} p_j \right)  ,
\end{equation}
where $K$ is an arbitrary constant with dimension energy per angular frequency. With this definition, the Hamiltonian becomes
\begin{equation}
\label{eq:ElmagH3}
H_\mathrm{cav} = \sum_j K \omega_j a^\ast_j a_j \ .
\end{equation}
In other words, the dimensionless number $a^\ast_j a_j = |a_j|^2$ gives the electromagnetic energy in mode $j$ in units of $K \omega_j$. Since we may choose the value of $K$ as we please, we could for example choose $K = \hbar$. This does not imply that we have quantized the field, but simply defines a scale for the coefficients $a_j$.

In absence of coupling to other systems, the equations of motion for the coefficients $a_j$ become
\begin{equation}
\label{eq:adyn}
\dot{a}_j = - \{H_\mathrm{cav},a_j\} = - i \omega_j a_j
\end{equation}
where $\{\cdot,\cdot\}$ is the Poisson bracket.

\subsection{Quantum versus classical field}

The transition from classical physics comes not from the choice of the constant $K$, but from imposing commutation relations on the coefficients $a_j, a_j^\ast \rightarrow a_j, a_j^\dagger$, such that 
\begin{equation}
\label{eq:aDefQuantum}
[a_j , a_{j'}^\dagger] = \delta_{j,j'} 
\end{equation}
for the choice $K = \hbar$. This leads to the familiar quantization relation $[q_j,p_{j'}] = i \hbar \delta_{j,j'}$ for the "positions" $q_j$ and "momenta" $p_j$. The Hamiltonian then becomes
\begin{equation}
\label{eq:ElmagH4}
H_\mathrm{cav} = \sum_j \hbar \omega_j \left(a^\dagger_j a_j + \frac{1}{2}\right),
\end{equation}
and $a_j^\dagger a_j$ is interpreted as the operator associated with the number of photons in mode $j$. The equation of motion is in this case given by the Heisenberg equation 
\begin{equation}
\label{eq:adynQuantum}
\dot{a}_j = \frac{i}{\hbar} [H_\mathrm{cav},a_j] = - i \omega_j a_j
\end{equation}
which is identical to the classical equation of motion in \eqref{eq:adyn}.

To be able to compare classical and quantum theories, it is convenient to simply choose $K = \hbar$ in both cases.

\section{Relating the electric field at the detector to the cavity output mode}
\label{app:Edet}

We have
\begin{align}
\label{eq:EoutputPlus}
 E^{(+)}(\mathbf{r},t)  &  
= \sum_j \sqrt{\frac{\hbar \omega_j}{2\varepsilon_0}}  e^{-i \omega_d t} b_j(t) w_j(\mathbf{r})  
\end{align}
and $E^{(-)}(\mathbf{r},t) = (E^{(+)}(\mathbf{r},t))^\dagger$. Here, the mode coefficients $b_j(t)$ are in the frame rotating at the frequency $\omega_d$. The mode functions $w_j(\mathbf{r})$ need not be specified, but we note that they satisfy the orthonormality condition
\begin{equation}
\label{eq:NormCond}
\int_V d\mathbf{r}  \, w^\ast_j(\mathbf{r})  w_{j'}(\mathbf{r}) = \delta_{j,j'}.
\end{equation}
Here, $V$ is the quantization volume. In the continuum limit, $V$ is infinitely large and the sums over $j$ turn into integrals. 

In the classical detector model, we are interested in the electric field squared averaged over the size of the detector, i.e., we need
\begin{align}
\label{eq:AveDetector}
& \int_{V} d\mathbf{r} \, f(\mathbf{r}) E^2(\mathbf{r},t) = \int_{V} f(\mathbf{r}) \sum_{j,j'} \frac{\hbar}{2\varepsilon_0} \sqrt{\omega_j \omega_{j'}} \\
& \times \left(b_j(t) b^\dagger_{j'}(t) w_j(\mathbf{r})  w^\ast_{j'}(\mathbf{r}) + b^\dagger_j(t) b_{j'}(t) w^\ast_j(\mathbf{r})  w_{j'}(\mathbf{r}) \right) \notag
\end{align}
where the function $f(\mathbf{r})$ has support only in the active region of the detector. We have neglected terms that oscillate at $\pm 2 \omega_d$, assuming that the detector does not react to such high frequency oscillations. The integral
\begin{equation}
\label{eq:Wintegral}
W_{j,j'} = \int_{V} d\mathbf{r} \, f(\mathbf{r}) w^\ast_j(\mathbf{r})   w_{j'}(\mathbf{r}) 
\end{equation}
depends on the phase difference accumulated by the two mode functions over the length $d$ of the detector in the direction of propagation. The mode functions have a wavelength $\lambda_j$ and we define the associated wave numbers $k_j = 2\pi/\lambda_j$. In the limit
\begin{equation}
\label{eq:LimitGood}
(k_j - k_{j'}) d \ll 2 \pi , 
\end{equation}
the accumulated phase difference is small, meaning that for all $j,j'$ that satisfy this, we have $W_{j,j'} \approx W$ independent of $j,j'$. In the opposite limit, the integral \eqref{eq:Wintegral} will give approximately a Kronecker delta as in Equation \eqref{eq:NormCond}. 

In our setup, we are only concerned with a frequency range of width $\sim \omega_\mathrm{if}$. This means that to be in the limit \eqref{eq:LimitGood}, we need $d \ll 2\pi c/\omega_\mathrm{if}$. As an example, the experiment in Ref.~\cite{Underwood2015PRA} used the intermediate frequency $\omega_\mathrm{if}/(2 \pi) = 80$ MHz, which gives $2\pi c/\omega_\mathrm{if} = 3.8$ m. This is clearly very well satisfied with standard photomultipliers. This means that we can replace the integral $W_{j,j'}$ with a constant $W$ in Equation \eqref{eq:AveDetector}. Focusing on a narrow frequency range means that we can also approximate $\sqrt{\omega_j \omega_{j'}}$ with $\omega_c$. This gives
\begin{align}
\label{eq:AveDetector2}
\int_{V} d\mathbf{r} \, f(\mathbf{r})  E^2(\mathbf{r},t) & \approx  \frac{\hbar \omega_c W}{2\varepsilon_0} \sum_{j,j'}  \left(b_j(t) b^\dagger_{j'}(t)  + b^\dagger_j(t) b_{j'}(t)  \right)  \\
& \equiv v^2 \left(a_\mathrm{det}(t) a^\dagger_\mathrm{det}(t) + a^\dagger_\mathrm{det}(t) a_\mathrm{det}(t) \right) \notag
\end{align}
with the constant
\begin{equation}
\label{eq:vDef}
v = \frac{\hbar \omega_c \pi D W}{\varepsilon_0} 
\end{equation}
and the detector mode defined as
\begin{equation}
\label{eq:aDetApp}
a_\mathrm{det}(t) = \frac{i}{\sqrt{2 \pi D}} \sum_j e^{-i \omega_j (t - t_1)} b_j(t_1) . 
\end{equation}

When normal ordering is required, one can in the same way show that  
\begin{align}
\label{eq:AveDetectorNorm}
& \int_{V} d\mathbf{r} \, f(\mathbf{r}) E^{(-)}(\mathbf{r},t)  E^{(+)}(\mathbf{r},t) \approx v^2 a^\dagger_\mathrm{det}(t) a_\mathrm{det}(t)
\end{align}
as long as we restrict ourselves to a narrow frequency range. Finally, for second order correlations, we have
\begin{align}
\label{eq:AveDetectorSecond}
& \int_{V} d\mathbf{r} \int_{V} d\mathbf{r'} \, f(\mathbf{r}) f(\mathbf{r'}) \\
& \times E^{(-)}(\mathbf{r},t)  E^{(-)}(\mathbf{r'},t + \tau) E^{(+)}(\mathbf{r'},t+\tau) E^{(+)}(\mathbf{r},t) \notag \\
& \approx v^4 a^\dagger_\mathrm{det}(t) a^\dagger_\mathrm{det}(t + \tau) a_\mathrm{det}(t + \tau) a_\mathrm{det}(t) .  \notag
\end{align}


\begin{thebibliography}{30}%
\makeatletter
\providecommand \@ifxundefined [1]{%
 \@ifx{#1\undefined}
}%
\providecommand \@ifnum [1]{%
 \ifnum #1\expandafter \@firstoftwo
 \else \expandafter \@secondoftwo
 \fi
}%
\providecommand \@ifx [1]{%
 \ifx #1\expandafter \@firstoftwo
 \else \expandafter \@secondoftwo
 \fi
}%
\providecommand \natexlab [1]{#1}%
\providecommand \enquote  [1]{``#1''}%
\providecommand \bibnamefont  [1]{#1}%
\providecommand \bibfnamefont [1]{#1}%
\providecommand \citenamefont [1]{#1}%
\providecommand \href@noop [0]{\@secondoftwo}%
\providecommand \href [0]{\begingroup \@sanitize@url \@href}%
\providecommand \@href[1]{\@@startlink{#1}\@@href}%
\providecommand \@@href[1]{\endgroup#1\@@endlink}%
\providecommand \@sanitize@url [0]{\catcode `\\12\catcode `\$12\catcode
  `\&12\catcode `\#12\catcode `\^12\catcode `\_12\catcode `\%12\relax}%
\providecommand \@@startlink[1]{}%
\providecommand \@@endlink[0]{}%
\providecommand \url  [0]{\begingroup\@sanitize@url \@url }%
\providecommand \@url [1]{\endgroup\@href {#1}{\urlprefix }}%
\providecommand \urlprefix  [0]{URL }%
\providecommand \Eprint [0]{\href }%
\providecommand \doibase [0]{http://dx.doi.org/}%
\providecommand \selectlanguage [0]{\@gobble}%
\providecommand \bibinfo  [0]{\@secondoftwo}%
\providecommand \bibfield  [0]{\@secondoftwo}%
\providecommand \translation [1]{[#1]}%
\providecommand \BibitemOpen [0]{}%
\providecommand \bibitemStop [0]{}%
\providecommand \bibitemNoStop [0]{.\EOS\space}%
\providecommand \EOS [0]{\spacefactor3000\relax}%
\providecommand \BibitemShut  [1]{\csname bibitem#1\endcsname}%
\let\auto@bib@innerbib\@empty
\bibitem [{\citenamefont {Aspelmeyer}\ \emph {et~al.}(2014)\citenamefont
  {Aspelmeyer}, \citenamefont {Kippenberg},\ and\ \citenamefont
  {Marquardt}}]{Aspelmeyer2014RMP}%
  \BibitemOpen
  \bibfield  {author} {\bibinfo {author} {\bibfnamefont {M.}~\bibnamefont
  {Aspelmeyer}}, \bibinfo {author} {\bibfnamefont {T.~J.}\ \bibnamefont
  {Kippenberg}}, \ and\ \bibinfo {author} {\bibfnamefont {F.}~\bibnamefont
  {Marquardt}},\ }\href@noop {} {\bibfield  {journal} {\bibinfo  {journal}
  {Rev. Mod. Phys.}\ }\textbf {\bibinfo {volume} {86}},\ \bibinfo {pages}
  {1391} (\bibinfo {year} {2014})}\BibitemShut {NoStop}%
\bibitem [{\citenamefont {Poot}\ and\ \citenamefont {van~der
  Zant}(2012)}]{Poot2012PhysRep}%
  \BibitemOpen
  \bibfield  {author} {\bibinfo {author} {\bibfnamefont {M.}~\bibnamefont
  {Poot}}\ and\ \bibinfo {author} {\bibfnamefont {H.~S.}\ \bibnamefont {van~der
  Zant}},\ }\href {\doibase http://dx.doi.org/10.1016/j.physrep.2011.12.004}
  {\bibfield  {journal} {\bibinfo  {journal} {Physics Reports}\ }\textbf
  {\bibinfo {volume} {511}},\ \bibinfo {pages} {273 } (\bibinfo {year}
  {2012})}\BibitemShut {NoStop}%
\bibitem [{\citenamefont {Metcalfe}(2014)}]{Metcalfe2014ApplPhysRev}%
  \BibitemOpen
  \bibfield  {author} {\bibinfo {author} {\bibfnamefont {M.}~\bibnamefont
  {Metcalfe}},\ }\href
  {http://scitation.aip.org/content/aip/journal/apr2/1/3/10.1063/1.4896029}
  {\bibfield  {journal} {\bibinfo  {journal} {Applied Physics Reviews}\
  }\textbf {\bibinfo {volume} {1}},\ \bibinfo {eid} {031105} (\bibinfo {year}
  {2014})}\BibitemShut {NoStop}%
\bibitem [{\citenamefont {Adler}\ and\ \citenamefont
  {Bassi}(2009)}]{Adler2009Science}%
  \BibitemOpen
  \bibfield  {author} {\bibinfo {author} {\bibfnamefont {S.~L.}\ \bibnamefont
  {Adler}}\ and\ \bibinfo {author} {\bibfnamefont {A.}~\bibnamefont {Bassi}},\
  }\href {\doibase 10.1126/science.1176858} {\bibfield  {journal} {\bibinfo
  {journal} {Science}\ }\textbf {\bibinfo {volume} {325}},\ \bibinfo {pages}
  {275} (\bibinfo {year} {2009})}\BibitemShut {NoStop}%
\bibitem [{\citenamefont {Arndt}\ and\ \citenamefont
  {Hornberger}(2014)}]{Arndt2014NatPhys}%
  \BibitemOpen
  \bibfield  {author} {\bibinfo {author} {\bibfnamefont {M.}~\bibnamefont
  {Arndt}}\ and\ \bibinfo {author} {\bibfnamefont {K.}~\bibnamefont
  {Hornberger}},\ }\href@noop {} {\bibfield  {journal} {\bibinfo  {journal}
  {Nat. Phys}\ }\textbf {\bibinfo {volume} {10}},\ \bibinfo {pages} {271}
  (\bibinfo {year} {2014})}\BibitemShut {NoStop}%
\bibitem [{\citenamefont {Diedrich}\ \emph {et~al.}(1989)\citenamefont
  {Diedrich}, \citenamefont {Bergquist}, \citenamefont {Itano},\ and\
  \citenamefont {Wineland}}]{Diedrich1989PRL}%
  \BibitemOpen
  \bibfield  {author} {\bibinfo {author} {\bibfnamefont {F.}~\bibnamefont
  {Diedrich}}, \bibinfo {author} {\bibfnamefont {J.~C.}\ \bibnamefont
  {Bergquist}}, \bibinfo {author} {\bibfnamefont {W.~M.}\ \bibnamefont
  {Itano}}, \ and\ \bibinfo {author} {\bibfnamefont {D.~J.}\ \bibnamefont
  {Wineland}},\ }\href {\doibase 10.1103/PhysRevLett.62.403} {\bibfield
  {journal} {\bibinfo  {journal} {Phys. Rev. Lett.}\ }\textbf {\bibinfo
  {volume} {62}},\ \bibinfo {pages} {403} (\bibinfo {year} {1989})}\BibitemShut
  {NoStop}%
\bibitem [{\citenamefont {Teufel}\ \emph {et~al.}(2011)\citenamefont {Teufel},
  \citenamefont {Donner}, \citenamefont {Li}, \citenamefont {Harlow},
  \citenamefont {Allman}, \citenamefont {Cicak}, \citenamefont {Sirois},
  \citenamefont {Whittaker}, \citenamefont {Lehnert},\ and\ \citenamefont
  {Simmonds}}]{Teufel2011Nature}%
  \BibitemOpen
  \bibfield  {author} {\bibinfo {author} {\bibfnamefont {J.~D.}\ \bibnamefont
  {Teufel}}, \bibinfo {author} {\bibfnamefont {T.}~\bibnamefont {Donner}},
  \bibinfo {author} {\bibfnamefont {D.}~\bibnamefont {Li}}, \bibinfo {author}
  {\bibfnamefont {J.~W.}\ \bibnamefont {Harlow}}, \bibinfo {author}
  {\bibfnamefont {M.~S.}\ \bibnamefont {Allman}}, \bibinfo {author}
  {\bibfnamefont {K.}~\bibnamefont {Cicak}}, \bibinfo {author} {\bibfnamefont
  {A.~J.}\ \bibnamefont {Sirois}}, \bibinfo {author} {\bibfnamefont {J.~D.}\
  \bibnamefont {Whittaker}}, \bibinfo {author} {\bibfnamefont {K.~W.}\
  \bibnamefont {Lehnert}}, \ and\ \bibinfo {author} {\bibfnamefont {R.~W.}\
  \bibnamefont {Simmonds}},\ }\href@noop {} {\bibfield  {journal} {\bibinfo
  {journal} {Nature}\ }\textbf {\bibinfo {volume} {475}},\ \bibinfo {pages}
  {359} (\bibinfo {year} {2011})}\BibitemShut {NoStop}%
\bibitem [{\citenamefont {Chan}\ \emph {et~al.}(2011)\citenamefont {Chan},
  \citenamefont {Alegre}, \citenamefont {Safavi-Naeini}, \citenamefont {Hill},
  \citenamefont {Krause}, \citenamefont {Gr\"{o}blacher}, \citenamefont
  {Aspelmeyer},\ and\ \citenamefont {Painter}}]{Chan2011Nature}%
  \BibitemOpen
  \bibfield  {author} {\bibinfo {author} {\bibfnamefont {J.}~\bibnamefont
  {Chan}}, \bibinfo {author} {\bibfnamefont {T.~P.~M.}\ \bibnamefont {Alegre}},
  \bibinfo {author} {\bibfnamefont {A.~H.}\ \bibnamefont {Safavi-Naeini}},
  \bibinfo {author} {\bibfnamefont {J.~T.}\ \bibnamefont {Hill}}, \bibinfo
  {author} {\bibfnamefont {A.}~\bibnamefont {Krause}}, \bibinfo {author}
  {\bibfnamefont {S.}~\bibnamefont {Gr\"{o}blacher}}, \bibinfo {author}
  {\bibfnamefont {M.}~\bibnamefont {Aspelmeyer}}, \ and\ \bibinfo {author}
  {\bibfnamefont {O.}~\bibnamefont {Painter}},\ }\href@noop {} {\bibfield
  {journal} {\bibinfo  {journal} {Nature}\ }\textbf {\bibinfo {volume} {478}},\
  \bibinfo {pages} {89} (\bibinfo {year} {2011})}\BibitemShut {NoStop}%
\bibitem [{\citenamefont {Underwood}\ \emph {et~al.}(2015)\citenamefont
  {Underwood}, \citenamefont {Mason}, \citenamefont {Lee}, \citenamefont {Xu},
  \citenamefont {Jiang}, \citenamefont {Shkarin}, \citenamefont {B\o{}rkje},
  \citenamefont {Girvin},\ and\ \citenamefont {Harris}}]{Underwood2015PRA}%
  \BibitemOpen
  \bibfield  {author} {\bibinfo {author} {\bibfnamefont {M.}~\bibnamefont
  {Underwood}}, \bibinfo {author} {\bibfnamefont {D.}~\bibnamefont {Mason}},
  \bibinfo {author} {\bibfnamefont {D.}~\bibnamefont {Lee}}, \bibinfo {author}
  {\bibfnamefont {H.}~\bibnamefont {Xu}}, \bibinfo {author} {\bibfnamefont
  {L.}~\bibnamefont {Jiang}}, \bibinfo {author} {\bibfnamefont {A.~B.}\
  \bibnamefont {Shkarin}}, \bibinfo {author} {\bibfnamefont {K.}~\bibnamefont
  {B\o{}rkje}}, \bibinfo {author} {\bibfnamefont {S.~M.}\ \bibnamefont
  {Girvin}}, \ and\ \bibinfo {author} {\bibfnamefont {J.~G.~E.}\ \bibnamefont
  {Harris}},\ }\href {\doibase 10.1103/PhysRevA.92.061801} {\bibfield
  {journal} {\bibinfo  {journal} {Phys. Rev. A}\ }\textbf {\bibinfo {volume}
  {92}},\ \bibinfo {pages} {061801} (\bibinfo {year} {2015})}\BibitemShut
  {NoStop}%
\bibitem [{\citenamefont {Purdy}\ \emph {et~al.}(2015)\citenamefont {Purdy},
  \citenamefont {Yu}, \citenamefont {Kampel}, \citenamefont {Peterson},
  \citenamefont {Cicak}, \citenamefont {Simmonds},\ and\ \citenamefont
  {Regal}}]{Purdy2015PRA}%
  \BibitemOpen
  \bibfield  {author} {\bibinfo {author} {\bibfnamefont {T.~P.}\ \bibnamefont
  {Purdy}}, \bibinfo {author} {\bibfnamefont {P.-L.}\ \bibnamefont {Yu}},
  \bibinfo {author} {\bibfnamefont {N.~S.}\ \bibnamefont {Kampel}}, \bibinfo
  {author} {\bibfnamefont {R.~W.}\ \bibnamefont {Peterson}}, \bibinfo {author}
  {\bibfnamefont {K.}~\bibnamefont {Cicak}}, \bibinfo {author} {\bibfnamefont
  {R.~W.}\ \bibnamefont {Simmonds}}, \ and\ \bibinfo {author} {\bibfnamefont
  {C.~A.}\ \bibnamefont {Regal}},\ }\href {\doibase 10.1103/PhysRevA.92.031802}
  {\bibfield  {journal} {\bibinfo  {journal} {Phys. Rev. A}\ }\textbf {\bibinfo
  {volume} {92}},\ \bibinfo {pages} {031802} (\bibinfo {year}
  {2015})}\BibitemShut {NoStop}%
\bibitem [{\citenamefont {Peterson}\ \emph {et~al.}(2016)\citenamefont
  {Peterson}, \citenamefont {Purdy}, \citenamefont {Kampel}, \citenamefont
  {Andrews}, \citenamefont {Yu}, \citenamefont {Lehnert},\ and\ \citenamefont
  {Regal}}]{Peterson2016PRL}%
  \BibitemOpen
  \bibfield  {author} {\bibinfo {author} {\bibfnamefont {R.~W.}\ \bibnamefont
  {Peterson}}, \bibinfo {author} {\bibfnamefont {T.~P.}\ \bibnamefont {Purdy}},
  \bibinfo {author} {\bibfnamefont {N.~S.}\ \bibnamefont {Kampel}}, \bibinfo
  {author} {\bibfnamefont {R.~W.}\ \bibnamefont {Andrews}}, \bibinfo {author}
  {\bibfnamefont {P.-L.}\ \bibnamefont {Yu}}, \bibinfo {author} {\bibfnamefont
  {K.~W.}\ \bibnamefont {Lehnert}}, \ and\ \bibinfo {author} {\bibfnamefont
  {C.~A.}\ \bibnamefont {Regal}},\ }\href {\doibase
  10.1103/PhysRevLett.116.063601} {\bibfield  {journal} {\bibinfo  {journal}
  {Phys. Rev. Lett.}\ }\textbf {\bibinfo {volume} {116}},\ \bibinfo {pages}
  {063601} (\bibinfo {year} {2016})}\BibitemShut {NoStop}%
\bibitem [{\citenamefont {Marquardt}\ \emph {et~al.}(2007)\citenamefont
  {Marquardt}, \citenamefont {Chen}, \citenamefont {Clerk},\ and\ \citenamefont
  {Girvin}}]{Marquardt2007PRL}%
  \BibitemOpen
  \bibfield  {author} {\bibinfo {author} {\bibfnamefont {F.}~\bibnamefont
  {Marquardt}}, \bibinfo {author} {\bibfnamefont {J.~P.}\ \bibnamefont {Chen}},
  \bibinfo {author} {\bibfnamefont {A.~A.}\ \bibnamefont {Clerk}}, \ and\
  \bibinfo {author} {\bibfnamefont {S.~M.}\ \bibnamefont {Girvin}},\ }\href
  {\doibase 10.1103/PhysRevLett.99.093902} {\bibfield  {journal} {\bibinfo
  {journal} {Phys. Rev. Lett.}\ }\textbf {\bibinfo {volume} {99}},\ \bibinfo
  {pages} {093902} (\bibinfo {year} {2007})}\BibitemShut {NoStop}%
\bibitem [{\citenamefont {Wilson-Rae}\ \emph {et~al.}(2007)\citenamefont
  {Wilson-Rae}, \citenamefont {Nooshi}, \citenamefont {Zwerger},\ and\
  \citenamefont {Kippenberg}}]{Wilson-Rae2007PRL}%
  \BibitemOpen
  \bibfield  {author} {\bibinfo {author} {\bibfnamefont {I.}~\bibnamefont
  {Wilson-Rae}}, \bibinfo {author} {\bibfnamefont {N.}~\bibnamefont {Nooshi}},
  \bibinfo {author} {\bibfnamefont {W.}~\bibnamefont {Zwerger}}, \ and\
  \bibinfo {author} {\bibfnamefont {T.~J.}\ \bibnamefont {Kippenberg}},\ }\href
  {\doibase 10.1103/PhysRevLett.99.093901} {\bibfield  {journal} {\bibinfo
  {journal} {Phys. Rev. Lett.}\ }\textbf {\bibinfo {volume} {99}},\ \bibinfo
  {pages} {093901} (\bibinfo {year} {2007})}\BibitemShut {NoStop}%
\bibitem [{\citenamefont {Khalili}\ \emph {et~al.}(2012)\citenamefont
  {Khalili}, \citenamefont {Miao}, \citenamefont {Yang}, \citenamefont
  {Safavi-Naeini}, \citenamefont {Painter},\ and\ \citenamefont
  {Chen}}]{Khalili2012PRA}%
  \BibitemOpen
  \bibfield  {author} {\bibinfo {author} {\bibfnamefont {F.~Y.}\ \bibnamefont
  {Khalili}}, \bibinfo {author} {\bibfnamefont {H.}~\bibnamefont {Miao}},
  \bibinfo {author} {\bibfnamefont {H.}~\bibnamefont {Yang}}, \bibinfo {author}
  {\bibfnamefont {A.~H.}\ \bibnamefont {Safavi-Naeini}}, \bibinfo {author}
  {\bibfnamefont {O.}~\bibnamefont {Painter}}, \ and\ \bibinfo {author}
  {\bibfnamefont {Y.}~\bibnamefont {Chen}},\ }\href {\doibase
  10.1103/PhysRevA.86.033840} {\bibfield  {journal} {\bibinfo  {journal} {Phys.
  Rev. A}\ }\textbf {\bibinfo {volume} {86}},\ \bibinfo {pages} {033840}
  (\bibinfo {year} {2012})}\BibitemShut {NoStop}%
\bibitem [{\citenamefont {Weinstein}\ \emph {et~al.}(2014)\citenamefont
  {Weinstein}, \citenamefont {Lei}, \citenamefont {Wollman}, \citenamefont
  {Suh}, \citenamefont {Metelmann}, \citenamefont {Clerk},\ and\ \citenamefont
  {Schwab}}]{Weinstein2014PRX}%
  \BibitemOpen
  \bibfield  {author} {\bibinfo {author} {\bibfnamefont {A.~J.}\ \bibnamefont
  {Weinstein}}, \bibinfo {author} {\bibfnamefont {C.~U.}\ \bibnamefont {Lei}},
  \bibinfo {author} {\bibfnamefont {E.~E.}\ \bibnamefont {Wollman}}, \bibinfo
  {author} {\bibfnamefont {J.}~\bibnamefont {Suh}}, \bibinfo {author}
  {\bibfnamefont {A.}~\bibnamefont {Metelmann}}, \bibinfo {author}
  {\bibfnamefont {A.~A.}\ \bibnamefont {Clerk}}, \ and\ \bibinfo {author}
  {\bibfnamefont {K.~C.}\ \bibnamefont {Schwab}},\ }\href {\doibase
  10.1103/PhysRevX.4.041003} {\bibfield  {journal} {\bibinfo  {journal} {Phys.
  Rev. X}\ }\textbf {\bibinfo {volume} {4}},\ \bibinfo {pages} {041003}
  (\bibinfo {year} {2014})}\BibitemShut {NoStop}%
\bibitem [{\citenamefont {Jayich}\ \emph {et~al.}(2012)\citenamefont {Jayich},
  \citenamefont {Sankey}, \citenamefont {B{\o}rkje}, \citenamefont {Lee},
  \citenamefont {Yang}, \citenamefont {Underwood}, \citenamefont {Childress},
  \citenamefont {Petrenko}, \citenamefont {Girvin},\ and\ \citenamefont
  {Harris}}]{Jayich2012NJP}%
  \BibitemOpen
  \bibfield  {author} {\bibinfo {author} {\bibfnamefont {A.~M.}\ \bibnamefont
  {Jayich}}, \bibinfo {author} {\bibfnamefont {J.~C.}\ \bibnamefont {Sankey}},
  \bibinfo {author} {\bibfnamefont {K.}~\bibnamefont {B{\o}rkje}}, \bibinfo
  {author} {\bibfnamefont {D.}~\bibnamefont {Lee}}, \bibinfo {author}
  {\bibfnamefont {C.}~\bibnamefont {Yang}}, \bibinfo {author} {\bibfnamefont
  {M.}~\bibnamefont {Underwood}}, \bibinfo {author} {\bibfnamefont
  {L.}~\bibnamefont {Childress}}, \bibinfo {author} {\bibfnamefont
  {A.}~\bibnamefont {Petrenko}}, \bibinfo {author} {\bibfnamefont {S.~M.}\
  \bibnamefont {Girvin}}, \ and\ \bibinfo {author} {\bibfnamefont {J.~G.~E.}\
  \bibnamefont {Harris}},\ }\href
  {http://stacks.iop.org/1367-2630/14/i=11/a=115018} {\bibfield  {journal}
  {\bibinfo  {journal} {New Journal of Physics}\ }\textbf {\bibinfo {volume}
  {14}},\ \bibinfo {pages} {115018} (\bibinfo {year} {2012})}\BibitemShut
  {NoStop}%
\bibitem [{\citenamefont {Safavi-Naeini}\ \emph {et~al.}(2013)\citenamefont
  {Safavi-Naeini}, \citenamefont {Chan}, \citenamefont {Hill}, \citenamefont
  {Gr\"blacher}, \citenamefont {Miao}, \citenamefont {Chen}, \citenamefont
  {Aspelmeyer},\ and\ \citenamefont {Painter}}]{Safavi-Naeini2013NJP}%
  \BibitemOpen
  \bibfield  {author} {\bibinfo {author} {\bibfnamefont {A.~H.}\ \bibnamefont
  {Safavi-Naeini}}, \bibinfo {author} {\bibfnamefont {J.}~\bibnamefont {Chan}},
  \bibinfo {author} {\bibfnamefont {J.~T.}\ \bibnamefont {Hill}}, \bibinfo
  {author} {\bibfnamefont {S.}~\bibnamefont {Gr\"blacher}}, \bibinfo {author}
  {\bibfnamefont {H.}~\bibnamefont {Miao}}, \bibinfo {author} {\bibfnamefont
  {Y.}~\bibnamefont {Chen}}, \bibinfo {author} {\bibfnamefont {M.}~\bibnamefont
  {Aspelmeyer}}, \ and\ \bibinfo {author} {\bibfnamefont {O.}~\bibnamefont
  {Painter}},\ }\href {http://stacks.iop.org/1367-2630/15/i=3/a=035007}
  {\bibfield  {journal} {\bibinfo  {journal} {New Journal of Physics}\ }\textbf
  {\bibinfo {volume} {15}},\ \bibinfo {pages} {035007} (\bibinfo {year}
  {2013})}\BibitemShut {NoStop}%
\bibitem [{\citenamefont {Lecocq}\ \emph {et~al.}(2015)\citenamefont {Lecocq},
  \citenamefont {Teufel}, \citenamefont {Aumentado},\ and\ \citenamefont
  {Simmonds}}]{Lecocq2015NatPhys}%
  \BibitemOpen
  \bibfield  {author} {\bibinfo {author} {\bibfnamefont {F.}~\bibnamefont
  {Lecocq}}, \bibinfo {author} {\bibfnamefont {J.~D.}\ \bibnamefont {Teufel}},
  \bibinfo {author} {\bibfnamefont {J.}~\bibnamefont {Aumentado}}, \ and\
  \bibinfo {author} {\bibfnamefont {R.~W.}\ \bibnamefont {Simmonds}},\
  }\href@noop {} {\bibfield  {journal} {\bibinfo  {journal} {Nature Physics}\
  }\textbf {\bibinfo {volume} {11}},\ \bibinfo {pages} {635} (\bibinfo {year}
  {2015})}\BibitemShut {NoStop}%
\bibitem [{\citenamefont {Carmichael}(1987)}]{Carmichael1987JOptSocAmB}%
  \BibitemOpen
  \bibfield  {author} {\bibinfo {author} {\bibfnamefont {H.~J.}\ \bibnamefont
  {Carmichael}},\ }\href@noop {} {\bibfield  {journal} {\bibinfo  {journal} {J.
  Opt. Soc. Am. B}\ }\textbf {\bibinfo {volume} {4}},\ \bibinfo {pages} {1588}
  (\bibinfo {year} {1987})}\BibitemShut {NoStop}%
\bibitem [{\citenamefont {Milloni}(1984)}]{Milloni1984AmJPhys}%
  \BibitemOpen
  \bibfield  {author} {\bibinfo {author} {\bibfnamefont {P.~W.}\ \bibnamefont
  {Milloni}},\ }\href@noop {} {\bibfield  {journal} {\bibinfo  {journal} {Am.
  J. Phys.}\ }\textbf {\bibinfo {volume} {52}},\ \bibinfo {pages} {340}
  (\bibinfo {year} {1984})}\BibitemShut {NoStop}%
\bibitem [{\citenamefont {Brahms}\ \emph {et~al.}(2012)\citenamefont {Brahms},
  \citenamefont {Botter}, \citenamefont {Schreppler}, \citenamefont {Brooks},\
  and\ \citenamefont {Stamper-Kurn}}]{Brahms2012PRL}%
  \BibitemOpen
  \bibfield  {author} {\bibinfo {author} {\bibfnamefont {N.}~\bibnamefont
  {Brahms}}, \bibinfo {author} {\bibfnamefont {T.}~\bibnamefont {Botter}},
  \bibinfo {author} {\bibfnamefont {S.}~\bibnamefont {Schreppler}}, \bibinfo
  {author} {\bibfnamefont {D.~W.~C.}\ \bibnamefont {Brooks}}, \ and\ \bibinfo
  {author} {\bibfnamefont {D.~M.}\ \bibnamefont {Stamper-Kurn}},\ }\href
  {\doibase 10.1103/PhysRevLett.108.133601} {\bibfield  {journal} {\bibinfo
  {journal} {Phys. Rev. Lett.}\ }\textbf {\bibinfo {volume} {108}},\ \bibinfo
  {pages} {133601} (\bibinfo {year} {2012})}\BibitemShut {NoStop}%
\bibitem [{\citenamefont {Safavi-Naeini}\ \emph {et~al.}(2012)\citenamefont
  {Safavi-Naeini}, \citenamefont {Chan}, \citenamefont {Hill}, \citenamefont
  {Alegre}, \citenamefont {Krause},\ and\ \citenamefont
  {Painter}}]{Safavi-Naeini2012PRL}%
  \BibitemOpen
  \bibfield  {author} {\bibinfo {author} {\bibfnamefont {A.~H.}\ \bibnamefont
  {Safavi-Naeini}}, \bibinfo {author} {\bibfnamefont {J.}~\bibnamefont {Chan}},
  \bibinfo {author} {\bibfnamefont {J.~T.}\ \bibnamefont {Hill}}, \bibinfo
  {author} {\bibfnamefont {T.~P.~M.}\ \bibnamefont {Alegre}}, \bibinfo {author}
  {\bibfnamefont {A.}~\bibnamefont {Krause}}, \ and\ \bibinfo {author}
  {\bibfnamefont {O.}~\bibnamefont {Painter}},\ }\href {\doibase
  10.1103/PhysRevLett.108.033602} {\bibfield  {journal} {\bibinfo  {journal}
  {Phys. Rev. Lett.}\ }\textbf {\bibinfo {volume} {108}},\ \bibinfo {pages}
  {033602} (\bibinfo {year} {2012})}\BibitemShut {NoStop}%
\bibitem [{\citenamefont {Clerk}\ \emph {et~al.}(2010)\citenamefont {Clerk},
  \citenamefont {Devoret}, \citenamefont {Girvin}, \citenamefont {Marquardt},\
  and\ \citenamefont {Schoelkopf}}]{Clerk2010RMP}%
  \BibitemOpen
  \bibfield  {author} {\bibinfo {author} {\bibfnamefont {A.~A.}\ \bibnamefont
  {Clerk}}, \bibinfo {author} {\bibfnamefont {M.~H.}\ \bibnamefont {Devoret}},
  \bibinfo {author} {\bibfnamefont {S.~M.}\ \bibnamefont {Girvin}}, \bibinfo
  {author} {\bibfnamefont {F.}~\bibnamefont {Marquardt}}, \ and\ \bibinfo
  {author} {\bibfnamefont {R.~J.}\ \bibnamefont {Schoelkopf}},\ }\href
  {\doibase 10.1103/RevModPhys.82.1155} {\bibfield  {journal} {\bibinfo
  {journal} {Rev. Mod. Phys.}\ }\textbf {\bibinfo {volume} {82}},\ \bibinfo
  {pages} {1155} (\bibinfo {year} {2010})}\BibitemShut {NoStop}%
\bibitem [{\citenamefont {Gardiner}\ and\ \citenamefont
  {Collett}(1985)}]{Gardiner1985PRA}%
  \BibitemOpen
  \bibfield  {author} {\bibinfo {author} {\bibfnamefont {C.~W.}\ \bibnamefont
  {Gardiner}}\ and\ \bibinfo {author} {\bibfnamefont {M.~J.}\ \bibnamefont
  {Collett}},\ }\href {\doibase 10.1103/PhysRevA.31.3761} {\bibfield  {journal}
  {\bibinfo  {journal} {Phys. Rev. A}\ }\textbf {\bibinfo {volume} {31}},\
  \bibinfo {pages} {3761} (\bibinfo {year} {1985})}\BibitemShut {NoStop}%
\bibitem [{\citenamefont {Spekkens}(2008)}]{Spekkens2008PRL}%
  \BibitemOpen
  \bibfield  {author} {\bibinfo {author} {\bibfnamefont {R.~W.}\ \bibnamefont
  {Spekkens}},\ }\href {\doibase 10.1103/PhysRevLett.101.020401} {\bibfield
  {journal} {\bibinfo  {journal} {Phys. Rev. Lett.}\ }\textbf {\bibinfo
  {volume} {101}},\ \bibinfo {pages} {020401} (\bibinfo {year}
  {2008})}\BibitemShut {NoStop}%
\bibitem [{\citenamefont {Rivi\`ere}\ \emph {et~al.}(2011)\citenamefont
  {Rivi\`ere}, \citenamefont {Del\'eglise}, \citenamefont {Weis}, \citenamefont
  {Gavartin}, \citenamefont {Arcizet}, \citenamefont {Schliesser},\ and\
  \citenamefont {Kippenberg}}]{Riviere2011PRA}%
  \BibitemOpen
  \bibfield  {author} {\bibinfo {author} {\bibfnamefont {R.}~\bibnamefont
  {Rivi\`ere}}, \bibinfo {author} {\bibfnamefont {S.}~\bibnamefont
  {Del\'eglise}}, \bibinfo {author} {\bibfnamefont {S.}~\bibnamefont {Weis}},
  \bibinfo {author} {\bibfnamefont {E.}~\bibnamefont {Gavartin}}, \bibinfo
  {author} {\bibfnamefont {O.}~\bibnamefont {Arcizet}}, \bibinfo {author}
  {\bibfnamefont {A.}~\bibnamefont {Schliesser}}, \ and\ \bibinfo {author}
  {\bibfnamefont {T.~J.}\ \bibnamefont {Kippenberg}},\ }\href {\doibase
  10.1103/PhysRevA.83.063835} {\bibfield  {journal} {\bibinfo  {journal} {Phys.
  Rev. A}\ }\textbf {\bibinfo {volume} {83}},\ \bibinfo {pages} {063835}
  (\bibinfo {year} {2011})}\BibitemShut {NoStop}%
\bibitem [{\citenamefont {Glauber}(1963)}]{Glauber1963PR}%
  \BibitemOpen
  \bibfield  {author} {\bibinfo {author} {\bibfnamefont {R.~J.}\ \bibnamefont
  {Glauber}},\ }\href {\doibase 10.1103/PhysRev.130.2529} {\bibfield  {journal}
  {\bibinfo  {journal} {Phys. Rev.}\ }\textbf {\bibinfo {volume} {130}},\
  \bibinfo {pages} {2529} (\bibinfo {year} {1963})}\BibitemShut {NoStop}%
\bibitem [{\citenamefont {Caves}(1980)}]{Caves1980PRL}%
  \BibitemOpen
  \bibfield  {author} {\bibinfo {author} {\bibfnamefont {C.~M.}\ \bibnamefont
  {Caves}},\ }\href {\doibase 10.1103/PhysRevLett.45.75} {\bibfield  {journal}
  {\bibinfo  {journal} {Phys. Rev. Lett.}\ }\textbf {\bibinfo {volume} {45}},\
  \bibinfo {pages} {75} (\bibinfo {year} {1980})}\BibitemShut {NoStop}%
\bibitem [{\citenamefont {O'Connell}\ \emph {et~al.}(2010)\citenamefont
  {O'Connell}, \citenamefont {Hofheinz}, \citenamefont {Ansmann}, \citenamefont
  {Bialczak}, \citenamefont {Lenander}, \citenamefont {Lucero}, \citenamefont
  {Neeley}, \citenamefont {Sank}, \citenamefont {Wang}, \citenamefont {Weides},
  \citenamefont {Wenner}, \citenamefont {Martinis},\ and\ \citenamefont
  {Cleland}}]{OConnell2010Nature}%
  \BibitemOpen
  \bibfield  {author} {\bibinfo {author} {\bibfnamefont {A.~D.}\ \bibnamefont
  {O'Connell}}, \bibinfo {author} {\bibfnamefont {M.}~\bibnamefont {Hofheinz}},
  \bibinfo {author} {\bibfnamefont {M.}~\bibnamefont {Ansmann}}, \bibinfo
  {author} {\bibfnamefont {R.~C.}\ \bibnamefont {Bialczak}}, \bibinfo {author}
  {\bibfnamefont {M.}~\bibnamefont {Lenander}}, \bibinfo {author}
  {\bibfnamefont {E.}~\bibnamefont {Lucero}}, \bibinfo {author} {\bibfnamefont
  {M.}~\bibnamefont {Neeley}}, \bibinfo {author} {\bibfnamefont
  {D.}~\bibnamefont {Sank}}, \bibinfo {author} {\bibfnamefont {H.}~\bibnamefont
  {Wang}}, \bibinfo {author} {\bibfnamefont {M.}~\bibnamefont {Weides}},
  \bibinfo {author} {\bibfnamefont {J.}~\bibnamefont {Wenner}}, \bibinfo
  {author} {\bibfnamefont {J.~M.}\ \bibnamefont {Martinis}}, \ and\ \bibinfo
  {author} {\bibfnamefont {A.~N.}\ \bibnamefont {Cleland}},\ }\href@noop {}
  {\bibfield  {journal} {\bibinfo  {journal} {Nature}\ }\textbf {\bibinfo
  {volume} {464}},\ \bibinfo {pages} {697} (\bibinfo {year}
  {2010})}\BibitemShut {NoStop}%
\bibitem [{\citenamefont {Meenehan}\ \emph {et~al.}(2015)\citenamefont
  {Meenehan}, \citenamefont {Cohen}, \citenamefont {MacCabe}, \citenamefont
  {Marsili}, \citenamefont {Shaw},\ and\ \citenamefont
  {Painter}}]{Meenehan2015PRX}%
  \BibitemOpen
  \bibfield  {author} {\bibinfo {author} {\bibfnamefont {S.~M.}\ \bibnamefont
  {Meenehan}}, \bibinfo {author} {\bibfnamefont {J.~D.}\ \bibnamefont {Cohen}},
  \bibinfo {author} {\bibfnamefont {G.~S.}\ \bibnamefont {MacCabe}}, \bibinfo
  {author} {\bibfnamefont {F.}~\bibnamefont {Marsili}}, \bibinfo {author}
  {\bibfnamefont {M.~D.}\ \bibnamefont {Shaw}}, \ and\ \bibinfo {author}
  {\bibfnamefont {O.}~\bibnamefont {Painter}},\ }\href {\doibase
  10.1103/PhysRevX.5.041002} {\bibfield  {journal} {\bibinfo  {journal} {Phys.
  Rev. X}\ }\textbf {\bibinfo {volume} {5}},\ \bibinfo {pages} {041002}
  (\bibinfo {year} {2015})}\BibitemShut {NoStop}%
\end{thebibliography}
\end{document}